\documentclass[a4paper,fleqn,usenatbib]{mnras}

\usepackage{mathptmx}
\usepackage[T1]{fontenc}
\usepackage{ae,aecompl}
\usepackage{multirow}
\usepackage{tabu}
\usepackage{float}

\usepackage{graphicx}	% Including figure files
\usepackage{amsmath}	% Advanced maths commands
\usepackage{amssymb}	% Extra maths symbols
\title[Instabilities in models of $\kappa$\,Cassiopeiae]{\huge Instabilities and pulsations in models 
of the B-type supergiant $\kappa$\,Cassiopeiae (HD\,2905)}
\author[A. P. Yadav, S. Joshi and W. Glatzel]{
Abhay Pratap Yadav$^{1,2}$\thanks{E-mail: abhaypratapbhu@yahoo.com}, 
Santosh Joshi$^{3}$ and
Wolfgang Glatzel$^{4}$\thanks{E-mail: wglatze@astro.physik.uni-goettingen.de} 
\\
$^{1}$Department of Physics $\&$ Astronomy, National Institute of Technology, Rourkela - 769008, Odisha, India \\
$^{2}$Government Model College Shahpura, Dindori - 481990, Madhya Pradesh, India \\
$^{3}$Aryabhatta Research Institute of Observational Sciences, Manora Peak, Nainital - 263002, India\\ 
$^{4}$Institut f\"ur Astrophysik (IAG), Georg-August-Universit\"at G\"ottingen, Friedrich-Hund-Platz 1, D-37077 G\"ottingen, Germany\\
}

\date{Accepted XXX. Received YYY; in original form ZZZ}

\pubyear{2020}

\begin{document}
\label{firstpage}
\pagerange{\pageref{firstpage}--\pageref{lastpage}}
\maketitle

\begin{abstract}
For the B-type supergiant $\kappa$ Cassiopeiae (HD\,2905) variabilities with periods between several hours and a few days 
have been observed both photometrically and spectroscopically. 
A recent study of this star by \citet{ssd_2018} has revealed variability with a dominant period of 2.7 days. To understand
this variability, we present a linear non-adiabatic stability analysis with respect to radial perturbations for models 
of $\kappa$\,Cassiopeiae. Instabilities associated with the fundamental mode and the 
first overtone are identified for models with masses between 27 M$_{\odot}$ and 44 M$_{\odot}$.
For selected models, the instabilities are followed into the non-linear regime by numerical simulations. 
As a result, finite amplitude pulsations with periods between 3 and 1.8 days are found. 
The model with a mass of 34.5 M$_{\odot}$ exhibits a pulsation period of 2.7 days consistent with the observations.
In the non-linear regime, the instabilities may cause a substantial inflation of the envelope. 
\end{abstract}

\begin{keywords}
instabilities -- stars: massive -- stars: mass-loss -- stars: oscillations --  
stars: supergiants -- stars: winds, outflows
\end{keywords}

%%%%%%%%%%%%%%%%%%%%%%%%%%%%%%%%%%%%%%%%%%%%%%%%%%

%%%%%%%%%%%%%%%%% BODY OF PAPER %%%%%%%%%%%%%%%%%%

% ================ Section 1 ===============

\section{Introduction}

%This is a simple template for authors to write new MNRAS papers.
%See \texttt{mnras\_sample.tex} for a more complex example, and \texttt{mnras\_guide.tex}
%for a full user guide.

%All papers should start with an Introduction section, which sets the work
%in context, cites relevant earlier studies in the field by \citet{Others2013},
%and describes the problem the authors aim to solve \citep[e.g.][]{Author2012}.
In several B-type stars, photometric as well as spectroscopic variabilities have been observed 
\citep[see e.g.,][]{waelkens_1998, saio_2006, aerts_2010arti, balona_2011, saesen_2013, rivinius_2016}. Particularly, high quality 
observations taken from space based telescopes (e.g. MOST, CoRoT, Kepler, BRITE) have considerably enhanced our 
understanding of variabilities in B-type stars.
%=================================================================OLD========================================
 Variability together with episodes of enhanced mass-loss have also been observed in some 
 B-type supergiants \citep[see e.g.,][]{kraus_2015, haucke_2018}.   

$\kappa$ Cassiopeiae or HD\,2905 is a B-type supergiant situated in the constellation Cassiopeia. 
 The presence of an astrosphere around $\kappa$ Cassiopeiae was revealed by observations taken with the Infrared 
 Astronomical Satellite \citep{buren_1988}. Further high resolution infrared observations 
 taken with the Spitzer Space Telescope reveal that the astrosphere of $\kappa$ Cassiopeiae seems to have arcuate structure 
 with several cirrus-type filaments \citep[see][]{gvaramadze_2011, katushkina_2018}. 
 From the existence and structure of the astrosphere around $\kappa$ Cassiopeiae, \citet{katushkina_2018}
 concludes and suggests that this star might be a runaway star.

Similar to other B-type supergiants, $\kappa$ Cassiopeiae exhibits variabilities and mass-loss.
Using photometry with a 36 cm Cassegrain telescope, \citet{elst_1979} has reported a variability with a
period of 2.19 hours for this star.
However,
\citet{percy_1981} could not confirm this two hours variability. Rather he identified a variability with a
period of the order of 7 days.    
Contrary to the findings of \citet{percy_1981}, observations taken at the UP State Observatory (now known as ARIES, Nainital) 
as reported by \citet{badalia_1982} indicate rapid
variabilities with two periods of 1.7 and 1.4 hours, respectively. 
Using photometry data of the Hipparcos catalogue, \citet{koen_2002} have found 2675 new variable stars including the supergiant 
$\kappa$ Cassiopeiae. These authors claim a photometric variability with a period of 2.6 days in this star.
The supergiant $\kappa$ Cassiopeiae is also observed by the BRITE satellites. However, so far pulsations have not 
yet been found in the preliminary 
analysis of the collected data \citep{rybicka_2018}.
Based on 1141 high resolution stellar spectra taken in a time span of approximately 2900 days, \citet{ssd_2018} 
have recently 
reported the presence of variabilities with periods mainly in the range between 2.5 days and 10 days for 
$\kappa$ Cassiopeiae.
A dominant period of 2.7 days present both in the spectral lines and in the Hipparcos space photometry has been identified by these authors. 
Although \citet{ssd_2018} have suggested that the variabilities might be associated with gravity modes or motions caused by subsurface 
convection, the cause of 
these variabilities is not properly understood.

Thus, with the motivation to understand the dominant variability of 2.7 days in $\kappa$ Cassiopeiae, we shall present here a linear non-adiabatic stability analysis of 
models for $\kappa$ Cassiopeiae. The result of instabilities will then be determined by following them into the non-linear regime for selected unstable models 
using a fully conservative numerical scheme. The present paper falls into 
five sections where a description of the models used is given in section 2, and the linear stability analysis together with its 
results are discussed in section 3. The non-linear simulations and their results are presented in section 4. 
A discussion and conclusions follow (section 5).

% ================ Section 2 ===============
% ================ Section 2 ===============

\section{Models for $\kappa$\,Cassiopeiae}
\label{models}
\begin{figure}
\centering $
\Large
\begin{array}{c}

   \scalebox{0.68}{ \input{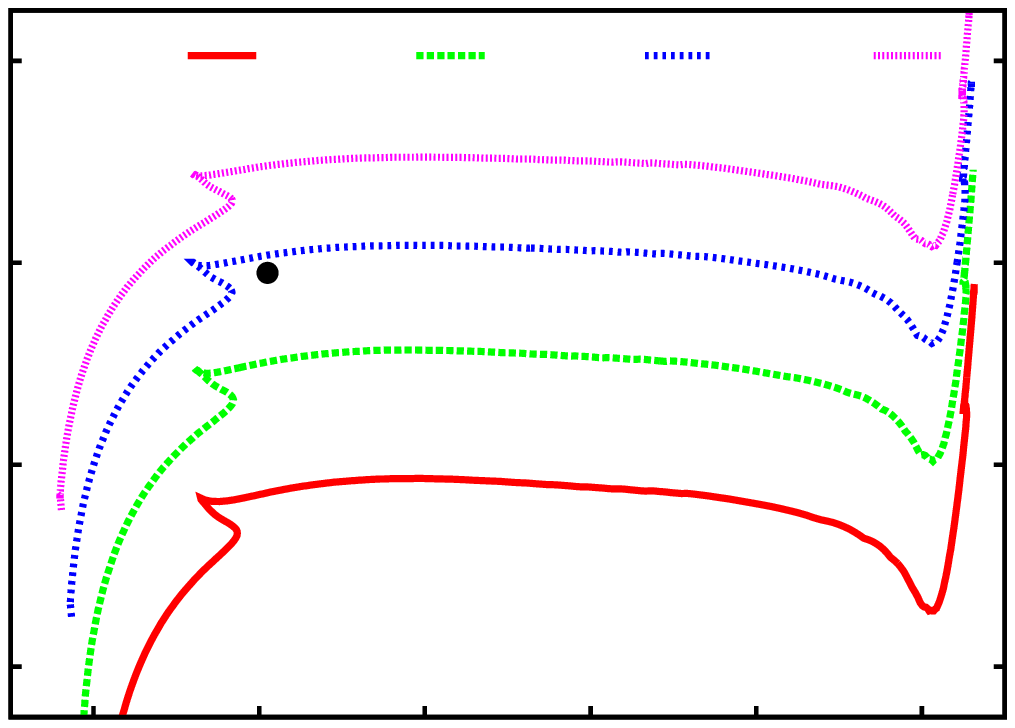} } 
   
 \end{array}$
 \caption{Evolutionary tracks of stars with solar chemical composition and having masses in the range between 30 and 45 M$_{\sun}$.  
The observed location of $\kappa$\,Cassiopeiae in the Hertzsprung-Russell diagram is marked by a thick dot.}
 \normalsize
 \label{track}
 \end{figure}

Although the star $\kappa$\,Cassiopeiae has been the subject of several studies \citep[see e.g.,][]{underhill_1979, percy_1981, hayes_1984, katushkina_2018},
its fundamental parameters (in particular its mass) are not 
precisely known. For the present investigation, we adopt the effective temperature (T$_{\rm{eff}}$ = 24600 K) and luminosity 
(log L/L$_{\sun}$ = 5.69) estimated by \citet{ssd_2018} using high resolution spectra of $\kappa$\,Cassiopeiae. These values are close to those 
determined by \citet{kudritzki_1999, smartt_2002} and \citet{evans_2004}. Due to the uncertainty of the mass, we consider 
a range of models with masses between 27 and 44 M$_{\sun}$. This mass range covers the mass of 33 M$_{\sun}$ suggested for $\kappa$\,Cassiopeiae by 
\citet{searle_2008} on the basis of stellar evolution calculations. Evolutionary tracks for models having solar chemical composition and masses of 30, 35, 40 and 45 M$_{\sun}$ respectively are shown in 
Fig. \ref{track} where a thick dot corresponds to the observed position of $\kappa$\,Cassiopeiae in the Hertzsprung-Russell (HR) diagram. These tracks have been 
generated by the `mad star' stellar evolution code\footnote[1]{\url{www.astro.wisc.edu/~townsend/static.php?ref=ez-web}}. The position of $\kappa$\,Cassiopeiae
in the HR diagram suggests that models representing this star have masses below and close to 40 M$_{\sun}$ (see Fig. \ref{track}).

The amplitudes of radial eigenmodes decay exponentially from the stellar surface to the core.
Thus the latter may be disregarded when studying stellar stability
\citep[see e.g.,][]{glatzel_1993, glatzel_1993b, saio_2011, yadav_2016, yadav_2017}. 
Accordingly our present study is restricted to envelope models of $\kappa$\,Cassiopeiae
thus disregarding any influence of nuclear energy generation and nucleosynthesis. 
The envelope models with solar chemical composition (X = 0.70, Y = 0.28 and Z = 0.02; where X, Y and Z represent the mass fraction of hydrogen, helium and 
heavy elements, respectively) are constructed by integrating the stellar structure equations as an initial value problem from the photosphere up to a sufficiently deep inner boundary 
(typically corresponding to a temperature of the order of 10$^{7}$ K). For the photosphere, Stefan-Boltzmann's law and a standard prescription 
for the photospheric pressure \citep[see e.g., section 11.2 of][]{kippenhahn_2012} are used as initial boundary conditions. Rotation as well as
magnetic fields are disregarded and OPAL opacity tables \citep{rogers_1992, rogers_1996, iglesias_1996} are used for the opacity. For the onset of 
convection, Schwarzschild's criterion is used and the convection is treated according to standard mixing length theory \citep{bohm_1958} with 1.5 pressure scale heights for the mixing length.

% ================ Section 3 ===============
% ================ Section 3 ===============

\section{Stability analysis}
To perform a linear stability analysis with respect to radial perturbations for models of $\kappa$\,Cassiopeiae with parameters as discussed in section 2, we have used the linearized 
perturbation equations given by \citet{gautschy_1990b}. This set of pulsation equations with four boundary conditions form a fourth order 
eigenvalue problem which is solved using the Riccati method in a similar way as described by \citet{gautschy_1990a}.
The solution of this 
system of equations leads to an infinite set of modes with complex eigenfrequencies 
($\sigma_{\rm{r}}$ + i$\sigma_{\rm{i}}$). For convenience, they will be normalized with the global free fall time 
($\sqrt{R^3/3GM}$; where $R$ is the 
stellar radius, $G$ denotes the gravitational constant and $M$ stands for the stellar mass) of the 
corresponding model. 
The real part of the 
eigenfrequency corresponds to the pulsation frequency, whereas the imaginary part provides information about damping or excitation 
of the mode. In the normalization adopted, negative imaginary parts ($\sigma_{\rm{i}}$ < 0) indicate excitation 
and instability while positive
imaginary parts
($\sigma_{\rm{i}}$ > 0) represent damping and stability. Interaction between pulsation and convection is still poorly understood. 
For simplicity, we have therefore used the `frozen in approximation' as introduced by \citet{baker_1965} in the 
present analysis. In this approximation, the Lagrangian
perturbation of the convective flux is assumed to vanish. In previous studies \citep{glatzel_1996, yadav_2017b} it was found to 
hold as long as energy transport is dominated by radiation diffusion.

\begin{figure*}
\centering $
\Large
\begin{array}{cc}

   \scalebox{0.69}{ \input{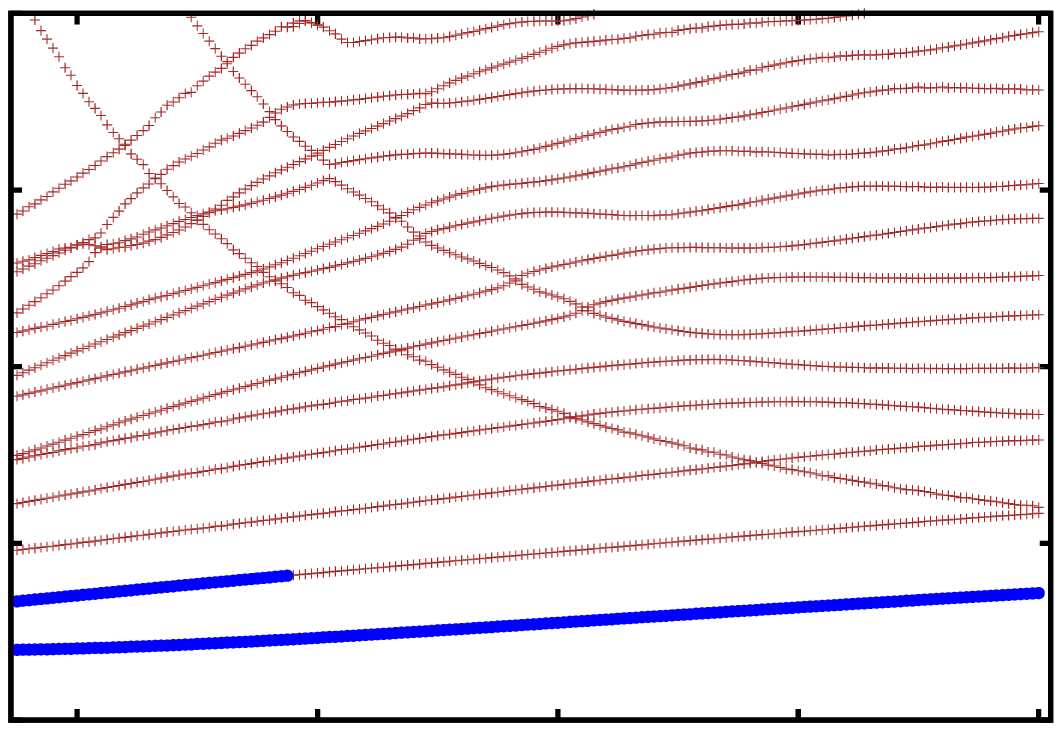} } 
   \scalebox{0.69}{ \input{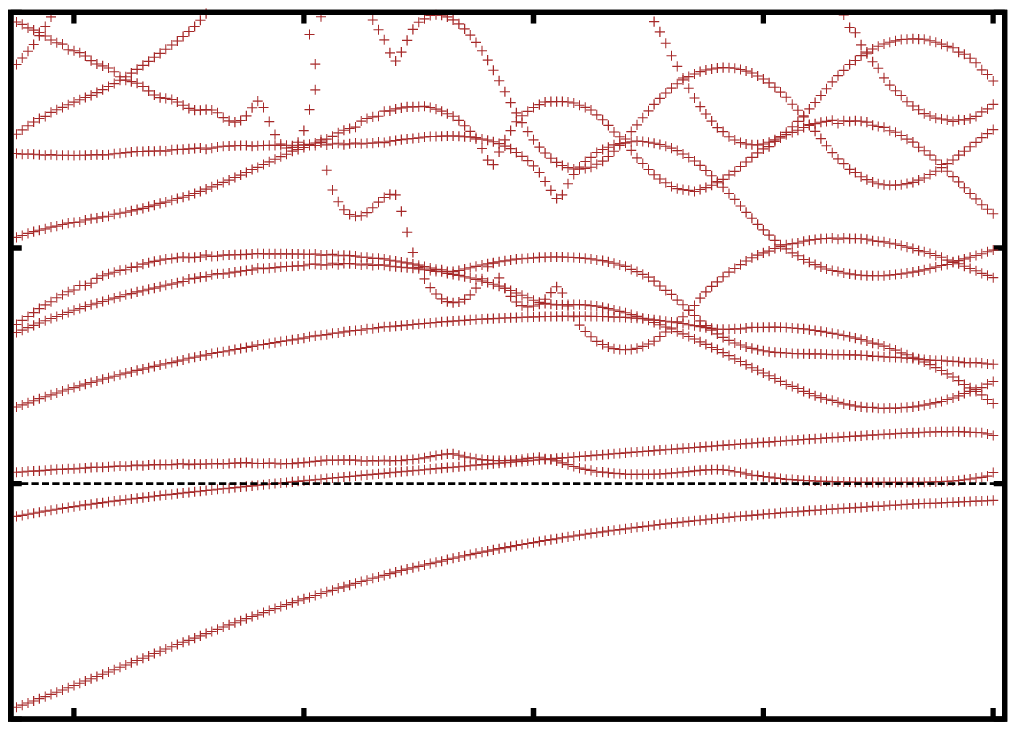} } \\
 \end{array}$
 \caption{Real (a) and imaginary (b) parts of the eigenfrequencies as a function of mass for models of 
 $\kappa$\,Cassiopeiae having solar chemical composition, an effective temperature 
 of T$_{\rm{eff}}$ = 24600 K and a luminosity of log L/L$_{\sun}$ = 5.69.  
 The eigenfrequencies are normalized with the global free fall timescale. 
 Thick blue lines in (a) and negative imaginary parts in (b) correspond to unstable modes. 
 For the outer boundary conditions, vanishing of the Lagrangian pressure perturbation and
 the validity of Stefan - Boltzmann's law have been adopted.}
 \normalsize
 \label{modal_1}
 \end{figure*}

\begin{figure*}
\centering $
\Large
\begin{array}{cc}

   \scalebox{0.69}{ \input{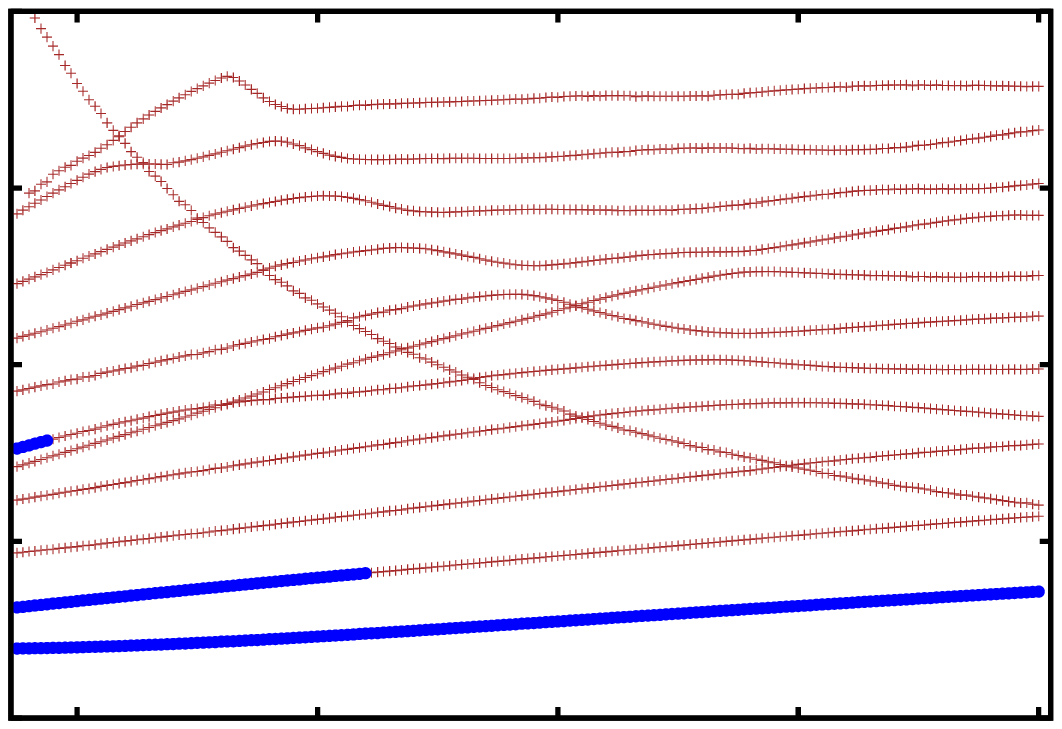} } 
   \scalebox{0.69}{ \input{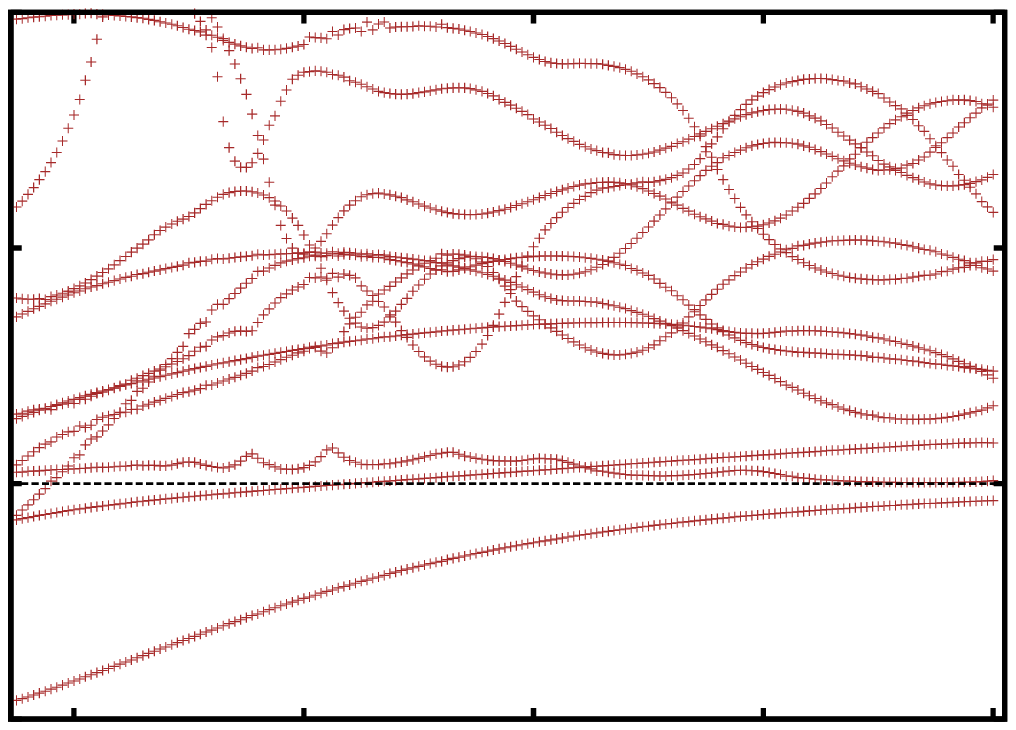} } \\
 \end{array}$
 \caption{Same as Fig. \ref{modal_1} but for boundary conditions consistent with those used in the 
 subsequent non-linear simulations.}
 \normalsize
 \label{modal_2}
 \end{figure*}

 \begin{figure}
\centering $
\Large
\begin{array}{c}
 \scalebox{0.62}{ \input{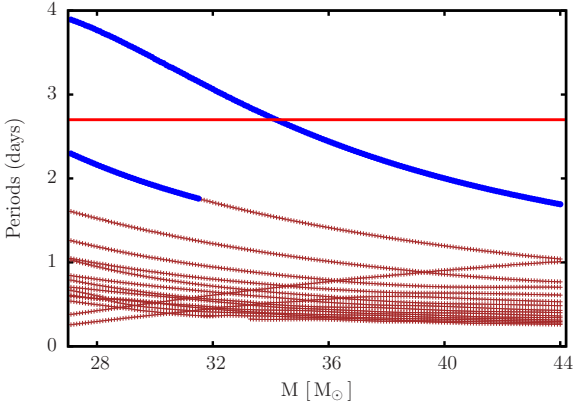} } \\
 \end{array}$
 \caption{Periods of various modes in models of $\kappa$\,Cassiopeiae. Thick blue lines correspond to unstable modes and the red line 
 denotes the observed dominant period of 2.7 days. Models in the mass range between 34 M$_{\odot}$ and 34.5 M$_{\odot}$ exhibit an unstable 
 fundamental mode with a period close to the observed value.}
 \normalsize
 \label{periods}
 \end{figure}

 \begin{figure}
\centering $
\Large
\begin{array}{cc}

  \scalebox{0.62}{ \input{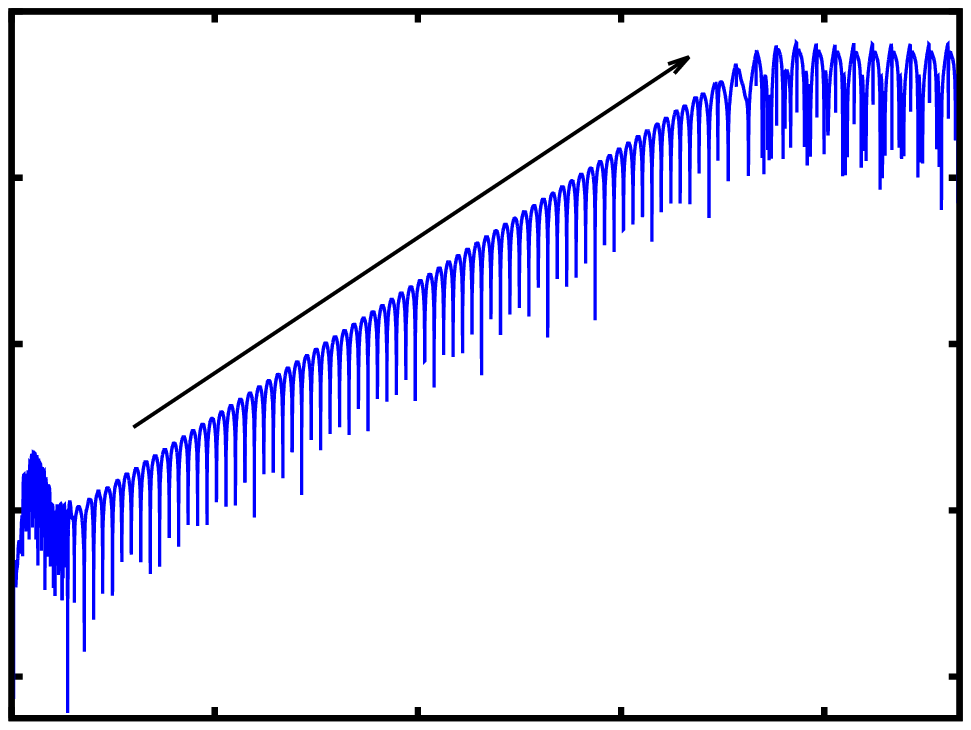} } \\
   \scalebox{0.62}{ \input{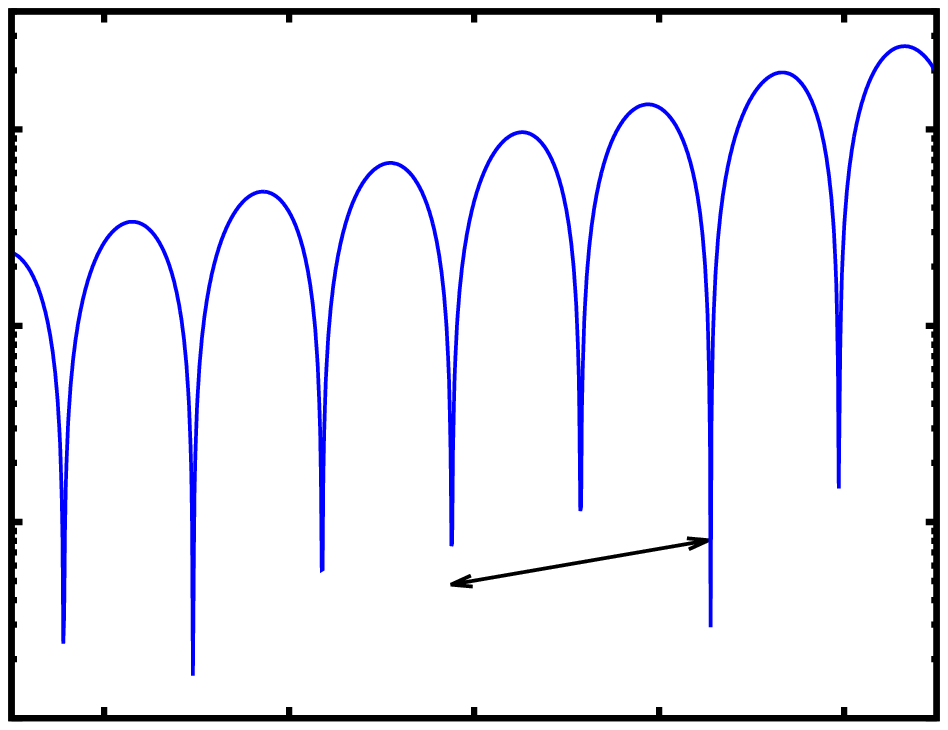} } \\
 \end{array}$
 \caption{Validation of the numerical scheme: For a numerical simulation the absolute velocity of the outermost gridpoint
is shown as a function time. The simulation starts from hydrostatic equilibrium, undergoes the linear phase
of exponential growth and finally ends in non-linear saturation. For the model considered, the linear 
stability analysis provides an 
unstable mode with a real part of 
 the eigenfrequency of $\sigma_{r}$ = 1.01 corresponding to a pulsation period of 2.72 days and an 
imaginary part of the 
 eigenfrequency of $\sigma_{i}$ = -0.11. In the linear phase, the growth rate observed in
 the numerical simulation in (a) is consistent with the value for $\sigma_{i}$ as determined by
the linear theory (slope of the arrow). Moreover, the pulsation period of 2.8 days deduced from (b) in the 
linear phase of the simulation is close to the value of
  2.72 days obtained from the linear analysis.}
 \normalsize
 \label{valid}
 \end{figure}

\begin{figure*}
\centering $
\LARGE
\begin{array}{ccc}
  \scalebox{0.455}{ \input{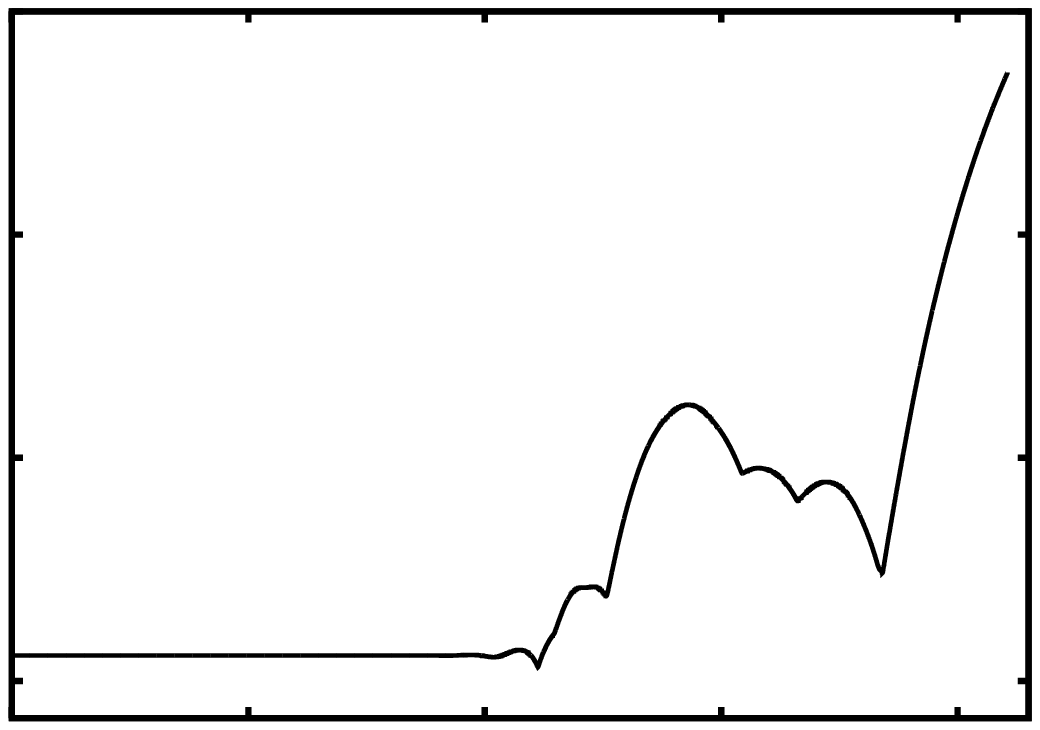} } 
   \scalebox{0.455}{ \input{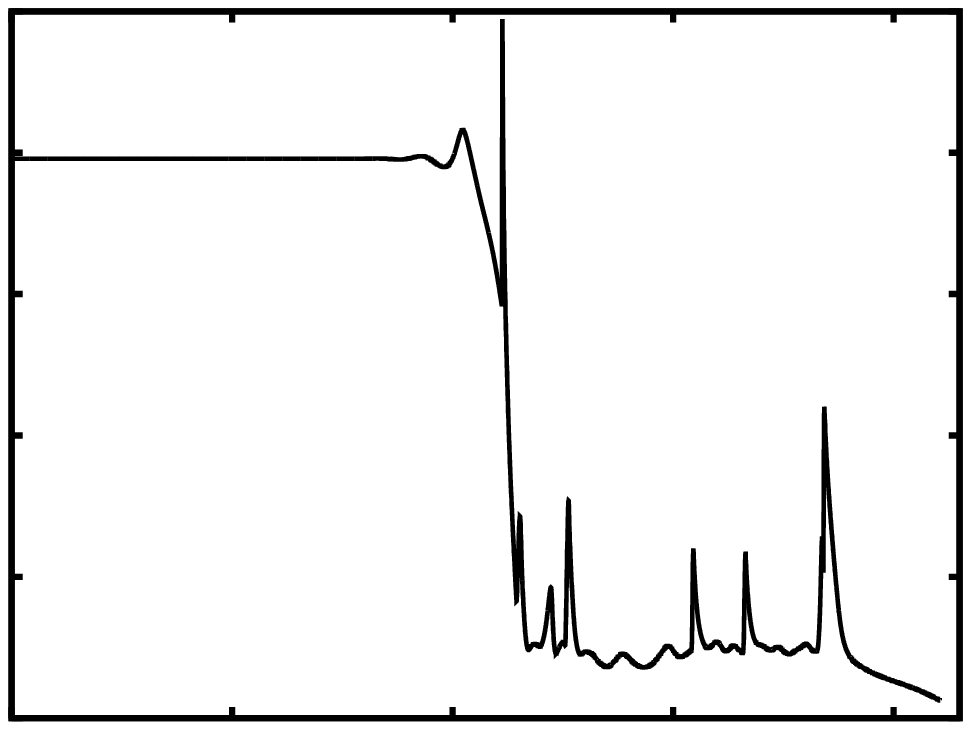} }
  \scalebox{0.455}{ \input{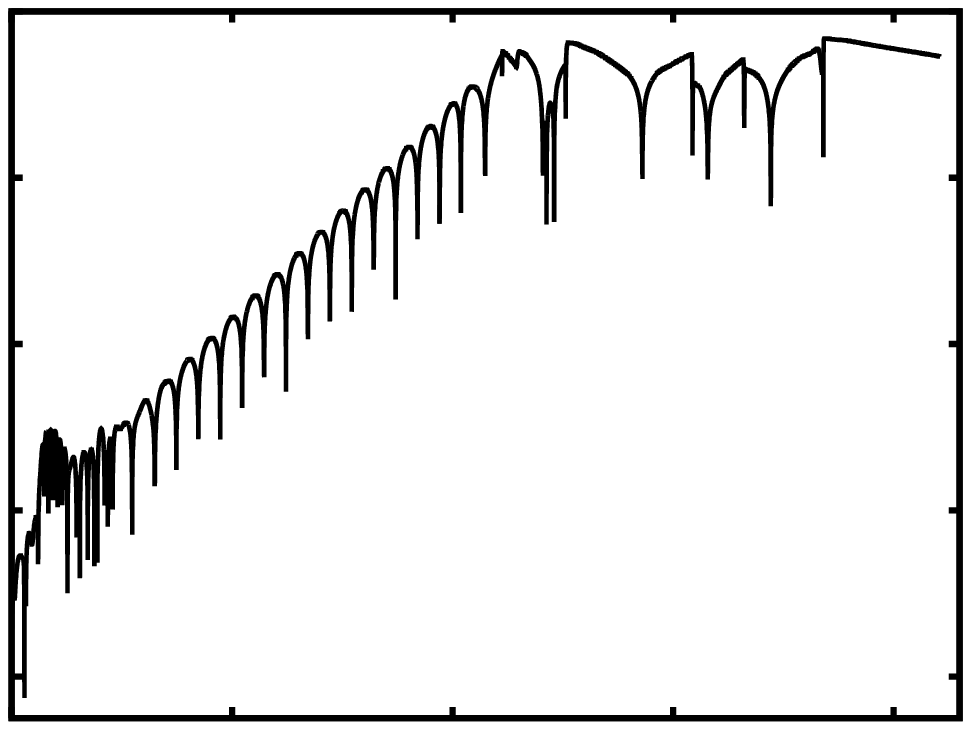} } \\
   \end{array}$
 \caption{Evolution of an instability into the non-linear regime for a model of $\kappa$-Cassiopeiae (HD\,2905) 
 having a mass of 27 M$_{\sun}$: Radius(a), temperature (b) and absolute velocity (c) of the outermost grid point 
 are given as a function of time. Note that in this case the instability leads to a substantial inflation of
 the model. As a consequence, the pulsation period is increased compared to the linearly determined value.}
 \normalsize 
 \label{27m_nonlin}
 \end{figure*}

 The results of the linear stability analysis are presented in terms of eigenfrequencies as a function 
 of stellar mass in Fig. \ref{modal_1}, where the real parts -- corresponding to the pulsation frequency -- 
 are shown in Fig. \ref{modal_1}(a) and the imaginary 
 parts -- indicating excitation or damping  -- are given in the Fig. \ref{modal_1}(b).
 Negative imaginary parts ($\sigma_{\rm{i}}$ < 0) correspond to excitation and instability.
 Real parts of the eigenfrequencies of excited modes are indicated by thick blue lines in 
 Fig. \ref{modal_1}(a). 
 From Fig. \ref{modal_1}, we deduce that all models considered with masses between 27 and 44 M$_{\sun}$ are unstable,
 where the instability affects the two lowest order modes (fundamental mode and first overtone). 
 Similar to previous studies \citep[e.g.,][]{yadav_2017b}, the growth rate and strength of the instabilities 
 increase with the luminosity to 
 mass ratio (see Fig. \ref{modal_1}b). For higher order modes, mode coupling phenomena 
 \citep[for more details see,][]{gautschy_1990b}, in particular avoided crossings can be 
 identified in the modal diagram given in Fig. \ref{modal_1}.

 %========================About the boundary conditions
 Although in many cases the results of the linear stability analysis are not substantially affected 
 by the choice of the outer boundary conditions, their influence for the models considered here
 needs to be discussed. The outer boundary conditions for the perturbation equations are ambiguous 
 because the outer boundary of the stellar model does not coincide with the physical outer boundary of the star.
 In fact for some stellar models, the choice of the outer boundary conditions was found to influence the 
 presence of instabilities originating from regions close to the stellar surface
 \citep[see e.g.][]{yadav_2016, yadav_2018}. 
 In order to check the dependence on the 
 outer boundary conditions of the results of the linear stability analysis for the models considered here, we 
 have therefore performed the linear stability analysis of our models 
 with different outer boundary conditions requiring zero heat storage and the gradient of compression to vanish there  
 \citep[for a detailed discussion of these boundary conditions see][]{grott_2005}. These outer boundary conditions are 
 used also for the subsequent numerical simulations following the instabilities into the non-linear 
 regime. The results of the linear stability analysis with these alternative 
 boundary conditions are given in Fig. \ref{modal_2}. Comparing Fig. \ref{modal_2} with Fig. \ref{modal_1}, we note 
 that for the stellar models considered the choice of the outer boundary conditions is of minor importance. 
 The unstable modes found in Fig. \ref{modal_1} are also present in Fig. \ref{modal_2}. Both the growth rates
 and the range of instabilities are almost identical for the different boundary conditions. However,
 an additional unstable mode is present in Fig. \ref{modal_2} for models with masses 
 below 28 M$_{\sun}$.    
 
 The pulsation periods of various modes are given as a function of mass in Fig.\,\ref{periods}. 
 Similar to Figs.\,\ref{modal_1} and \ref{modal_2}, 
 blue thick lines represent unstable modes. A red horizontal line corresponds to the period of 2.7 days
 observed in $\kappa$-Cassiopeiae. 
 The pulsation period of the unstable fundamental mode lies between 3.9 days and 1.6 days for the models considered 
 while the period of the unstable first overtone lies below 2.3 days. From Fig.\,\ref{periods}, we thus deduce 
 that models with masses between 34 and 34.5 M$_{\sun}$ provide (linear) pulsation periods close to the observed 
 period of 2.7 days.

 % ================ Section 4 ===============
 % ================ Section 4 ===============

\section{Instabilities in the non-linear regime}
The presence of at least two unstable modes in models of $\kappa$\,Cassiopeiae has been found on the basis of
a linear stability analysis. 
In order to determine the final fate of unstable models, the instabilities have to be followed into the non-linear regime.
As discussed by \citet{yadav_2017b} instabilities of stellar models may lead to finite amplitude 
pulsations, eruptions of surface layers or the rearrangement of the stellar structure. 
To follow the instabilities into the non-linear regime for selected unstable models 
of $\kappa$\,Cassiopeiae, we have used the numerical scheme described by 
\citet{grott_2005}.
This numerical scheme provides the extremely high precision necessary for the simulation
of finite amplitude pulsations and instabilities in the non-linear regime.
In particular, the energy balance is satisfied with extremely high precision. 
For the importance of a correct energy balance in the context of stellar pulsations we refer the reader 
to earlier studies 
\citep[see e.g.,][]{grott_2005, glatzel_2016, yadav_2017b}. Accordingly we have examined here the 
evolution of the error in the energy balance during the simulation of the evolution of an instability 
from hydrostatic equilibrium through the linear phase of exponential growth into the non-linear regime. 
In the non-linear regime, shock waves are expected to form. To handle the discontinuities 
caused by shocks, an artificial viscosity is introduced. It vanishes except close to a discontinuity. 
For details, the reader is 
referred to \citet{grott_2005} and 
\citet{yadav_2017b}. 

\subsection{Validation of the numerical scheme}

\label{validation}
 Apart from satisfying the energy balance with high precision, we require 
 the numerical scheme used for following instabilities into the non-linear 
 regime to represent the results of the independent linear stability analysis.
 For validation, we require the numerical code to start from a model
 in hydrostatic equilibrium and to pick up the physical instability 
 with the period and growth rate as predetermined independently by the linear 
 analysis from numerical noise without any external perturbation.
 To prove this property of the scheme, we present the 
 evolution of an instability from hydrostatic equilibrium through the linear phase of 
 exponential growth into the non-linear regime for a model with a mass of 34 M$_{\sun}$ in Fig \ref{valid}. 
 The linear stability analysis for this
 model of $\kappa$\,Cassiopeiae provides an unstable mode where the real part of the associated eigenfrequency 
 $\sigma_{r}$ = 1.01 corresponds to a pulsation period of 2.72 days. The imaginary part of the
 eigenfrequency $\sigma_{i}$ = -0.11 represents the growth rate of the mode. This period and growth rate 
 should appear in the linear phase of the the numerical simulation. As a result of the simulation the absolute velocity 
 of the outermost grid point is given as a function of time in Fig. \ref{valid}. The evolution of the instability 
 starts from hydrostatical equilibrium with a numerical noise of the order of 10$^{-4}$ cm s$^{-1}$. 
 Without any external perturbation the code picks up an oscillatory exponentially growing instability:
 This is the expected linear phase of exponential growth. Both the growth rate and the pulsation period
 of 2.8 days observed in the simulation are consistent with the eigenfrequency provided and predetermined
 by the linear stability analysis. 
 After the linear phase the velocity amplitude saturates and the 
 evolution finally enters the non-linear phase of finite amplitude pulsations.
 Thus the numerical code has been shown to pick up the correct physical instability without any 
 external manipulation. Additional numerical instabilities do not seem to be present.

%=============================

%=============================

%Due to the presence of artificial viscosity,

\begin{figure*}
\centering $
\LARGE
\begin{array}{ccccccccc}
  \scalebox{0.455}{ \input{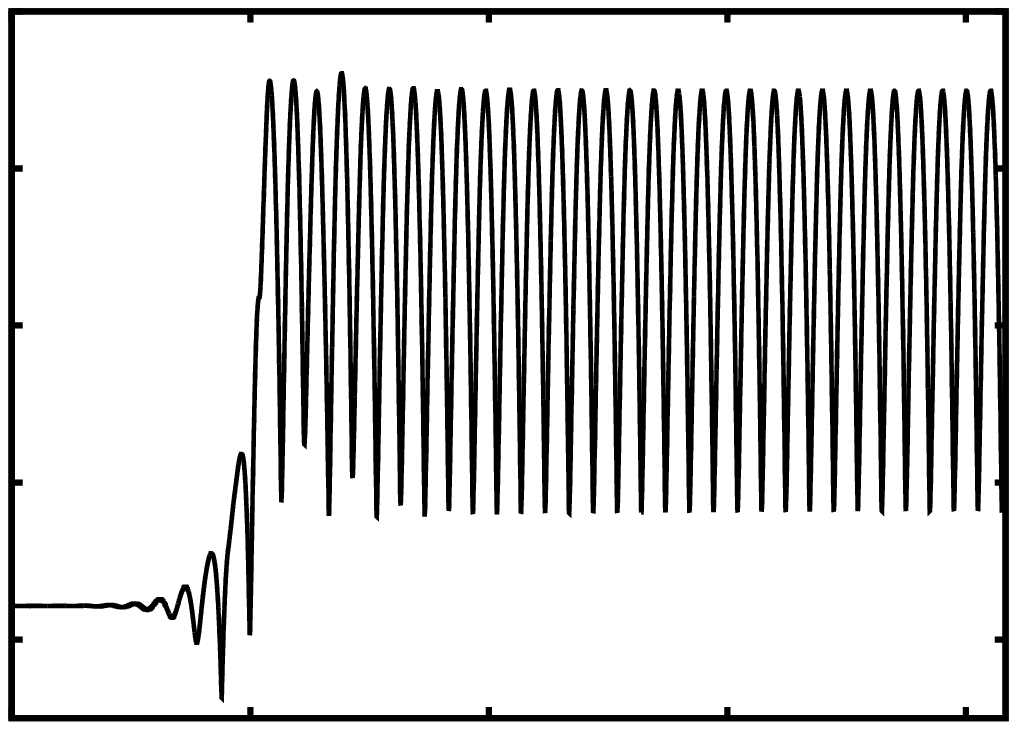} } 
  \scalebox{0.455}{ \input{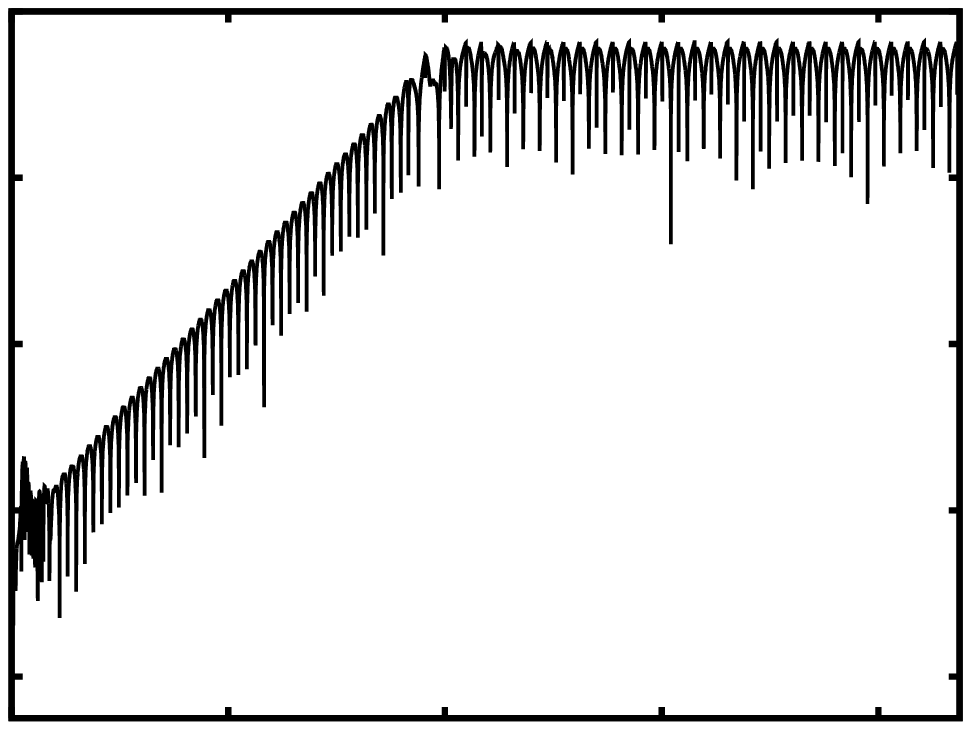} } \\
   \scalebox{0.455}{ \input{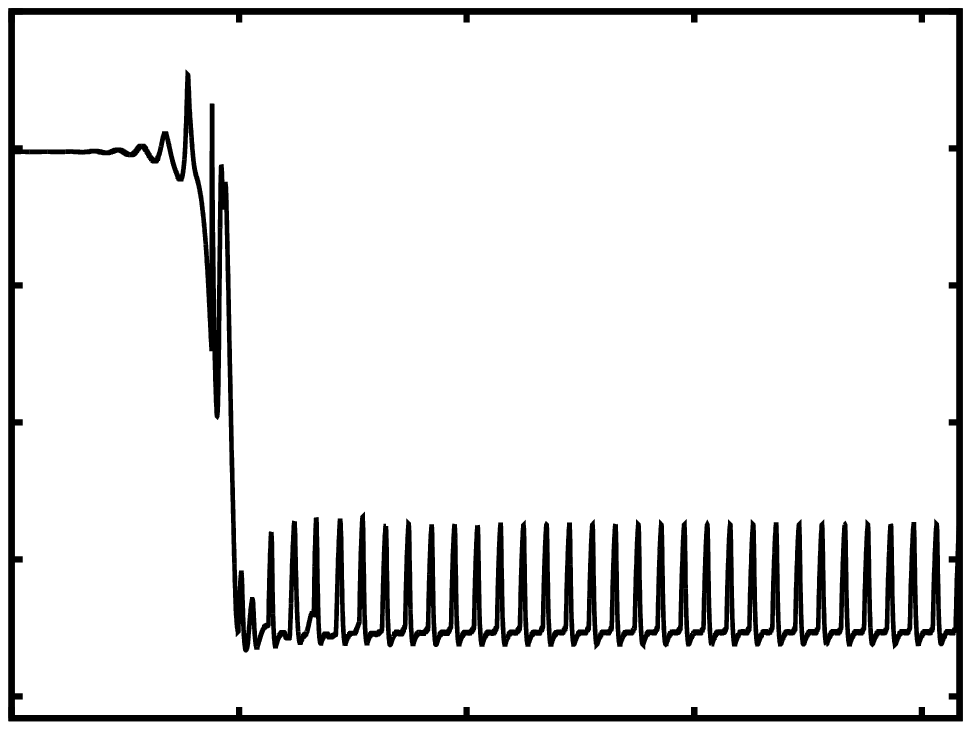} } 

%   \scalebox{0.455}{ \input{32m_den_I} } 
  \scalebox{0.455}{ \input{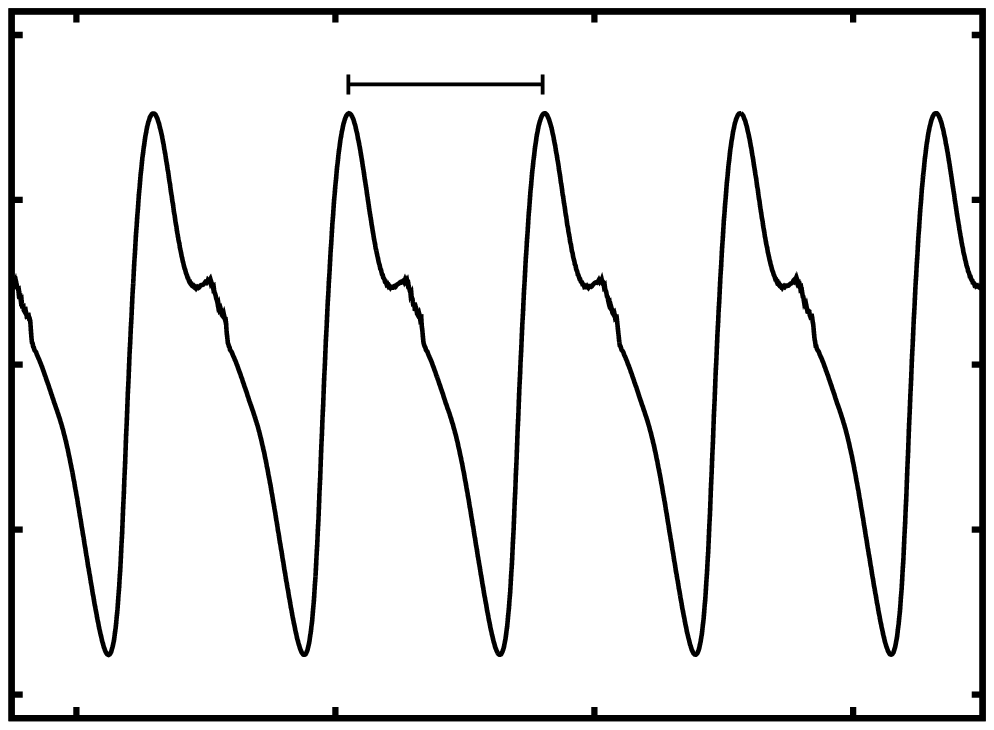} } \\
   
 %   \scalebox{0.455}{ \input{32m_kin_I} } \\
   
 %  \scalebox{0.455}{ \input{32m_ac_I} } 
 
 % \scalebox{0.455}{ \input{32m_energy_I} } 
  % \scalebox{0.455}{ \input{32m_error_I} } \\
   \end{array}$
 \caption{Evolution of an instability into the non-linear regime for a model of $\kappa$-Cassiopeiae (HD\,2905) with 
 a mass of 32 M$_{\sun}$: Radius (a), absolute velocity (b), 
 temperature (c) at the outermost grid point and variations of 
 the bolometric magnitude (d) are given as a function of time. 
 This model finally exhibits finite amplitude pulsations with a period of 3 days.}
 \normalsize 
 \label{32m_nonlin}
 \end{figure*}

 \begin{figure*}
\centering $
\LARGE
\begin{array}{ccc}
  \scalebox{0.455}{ \input{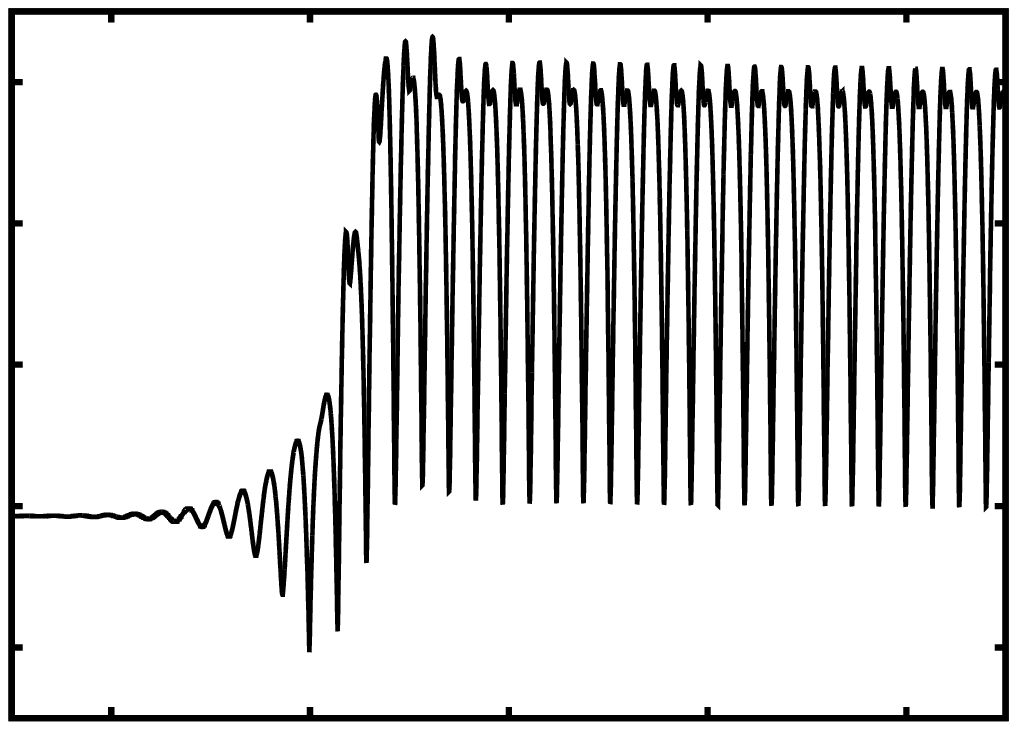} } 
  \scalebox{0.455}{ \input{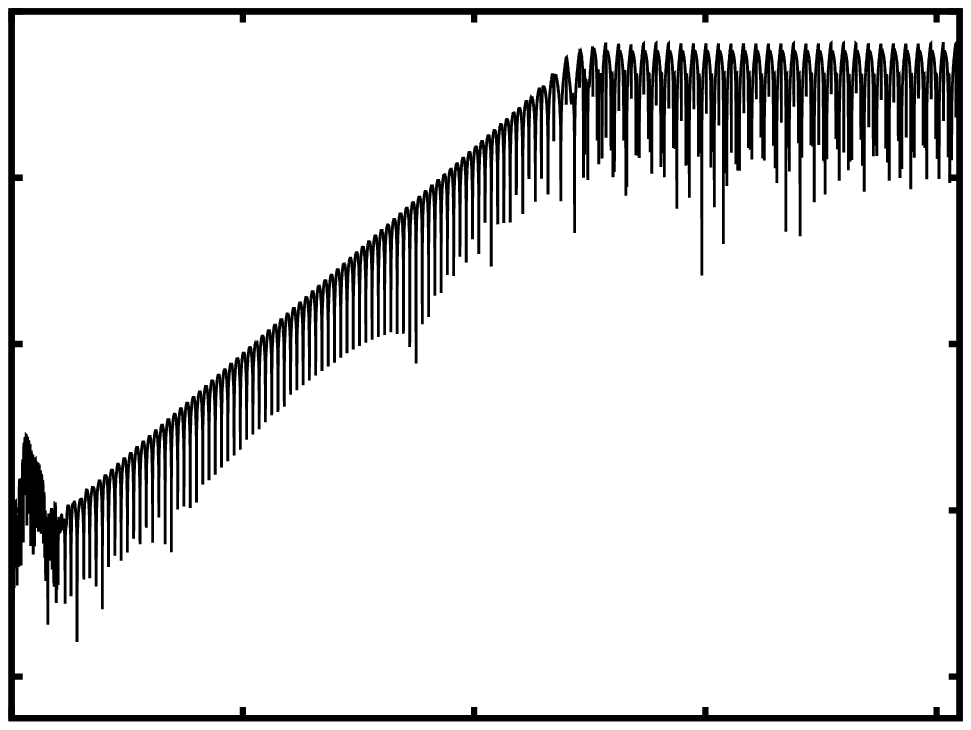} }
  % \scalebox{0.455}{ \input{345m_tem} } \\
% \scalebox{0.455}{ \input{345m_dens} } 
  \scalebox{0.455}{ \input{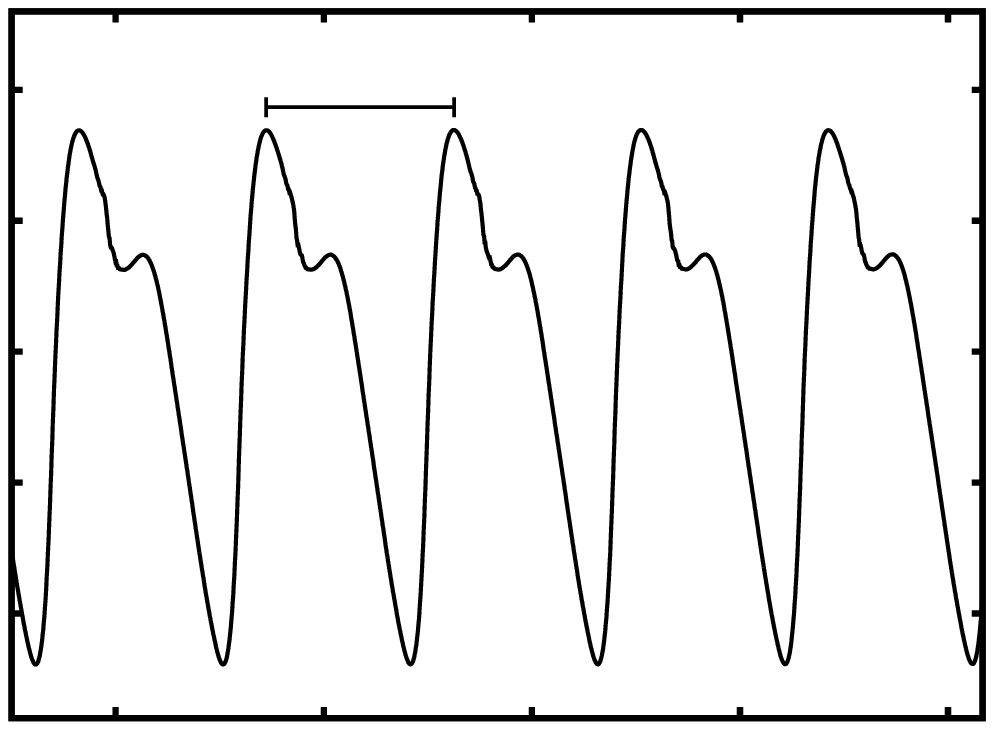} } \\
 %  \scalebox{0.455}{ \input{345m_kin} } \\
 %  \scalebox{0.455}{ \input{345m_acou} } 

  % \scalebox{0.455}{ \input{345m_energies} } 
 %  \scalebox{0.455}{ \input{345m_error} } \\
   \end{array}$
 \caption{Same as Fig.\,\ref{32m_nonlin} but for a model of $\kappa$-Cassiopeiae (HD\,2905) with a mass of 34.5 M$_{\sun}$: Radius (a),
 absolute velocity (b) at the outermost grid point and variations of the bolometric magnitude (c) are given as a function of time. 
 This model finally exhibits finite amplitude pulsations with a period of 2.7 days, 
 which exactly matches the observed dominant period of the star.}
 \normalsize 
 \label{345m_nonlin}
 \end{figure*}

 \begin{figure*}
\centering $
\LARGE
\begin{array}{cccc}
  \scalebox{0.455}{ \input{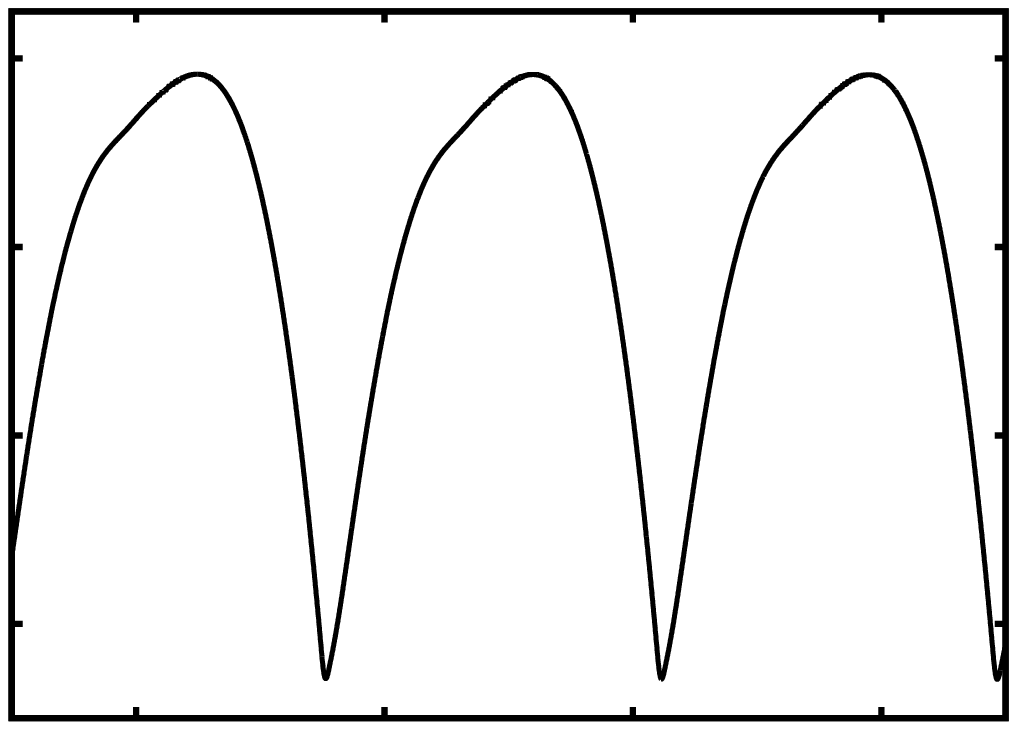} } 
  \scalebox{0.455}{ \input{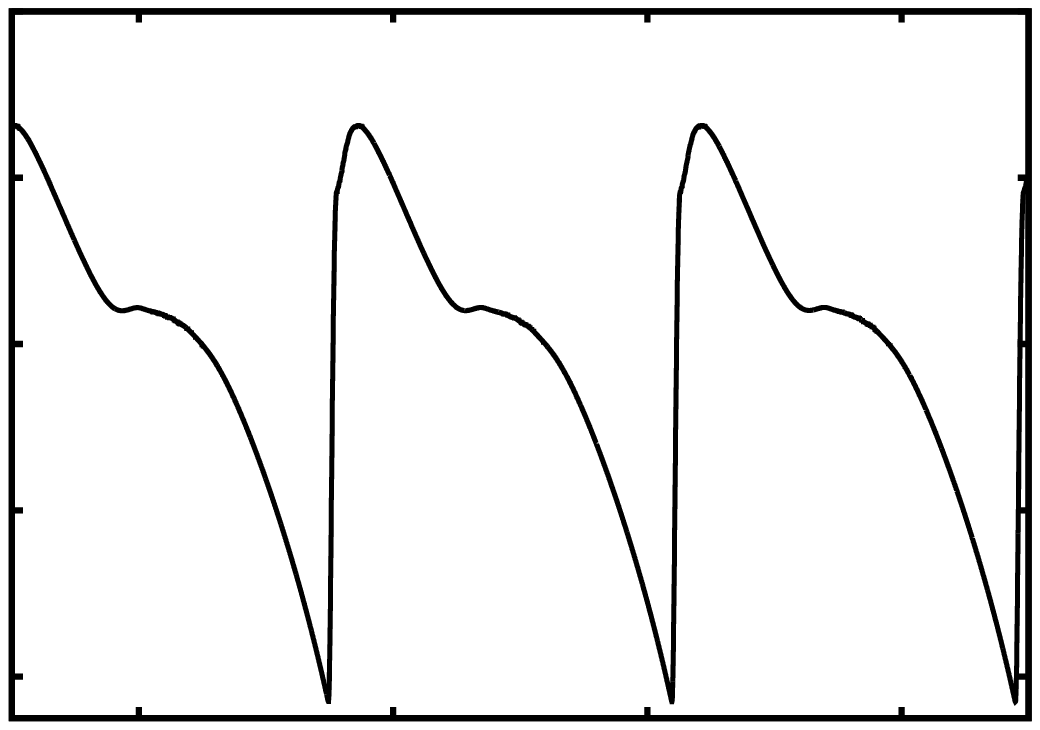} } \\
   \scalebox{0.455}{ \input{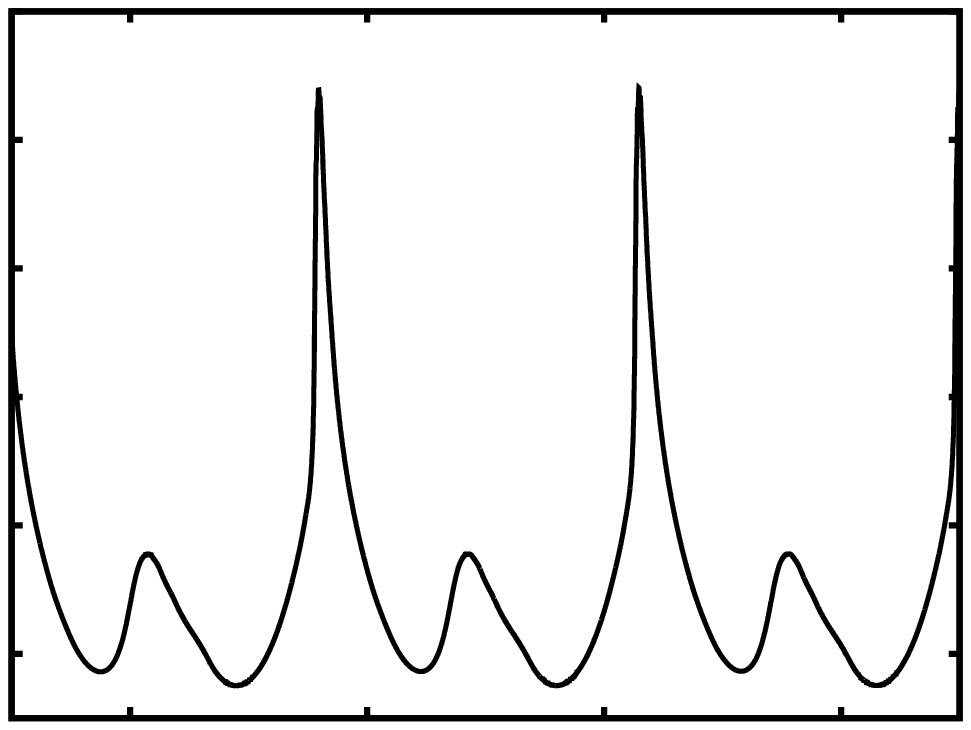} } 
 \scalebox{0.455}{ \input{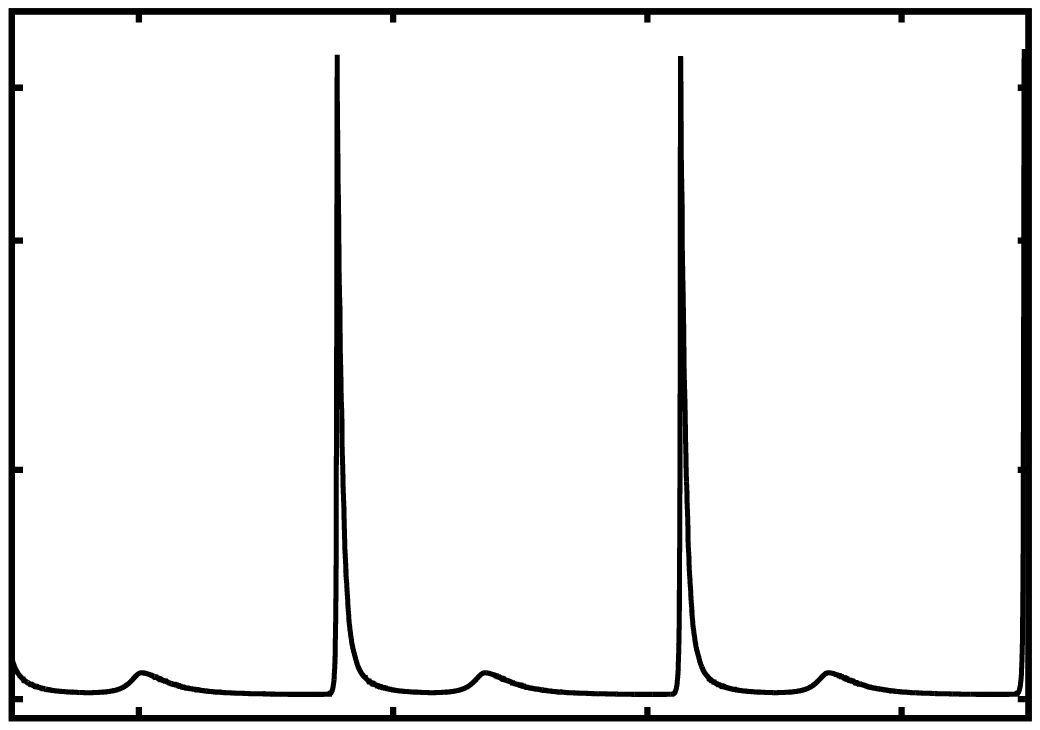} } 
 % \scalebox{0.455}{ \input{345m_PP} }\\
 %  \scalebox{0.455}{ \input{345m_kin} } \\
 %  \scalebox{0.455}{ \input{345m_acou} } 
%   \scalebox{0.455}{ \input{345m_energies} } 
 %  \scalebox{0.455}{ \input{345m_error} } \\
   \end{array}$
 \caption{ Finite amplitude pulsation with a period of 2.7 days in the model with a mass of 34.5 M$_{\sun}$. Variations in radius (a), absolute velocity (b), temperature (c) and 
 density (d) are given to a grid point close to the photosphere. }
 % Same as Fig.\,\ref{32m_nonlin} but for a model of $\kappa$-Cassiopeiae (HD\,2905) with a mass of 34.5 M$_{\sun}$. 
 %This model finally exhibits finite amplitude pulsations with a period of 2.7 days, 
 %which exactly matches the observed dominant period of the star.}
 \normalsize 
 \label{345m_nonlinP}
 \end{figure*}

 \begin{figure*}
\centering $
\LARGE
\begin{array}{ccc}
  \scalebox{0.455}{ \input{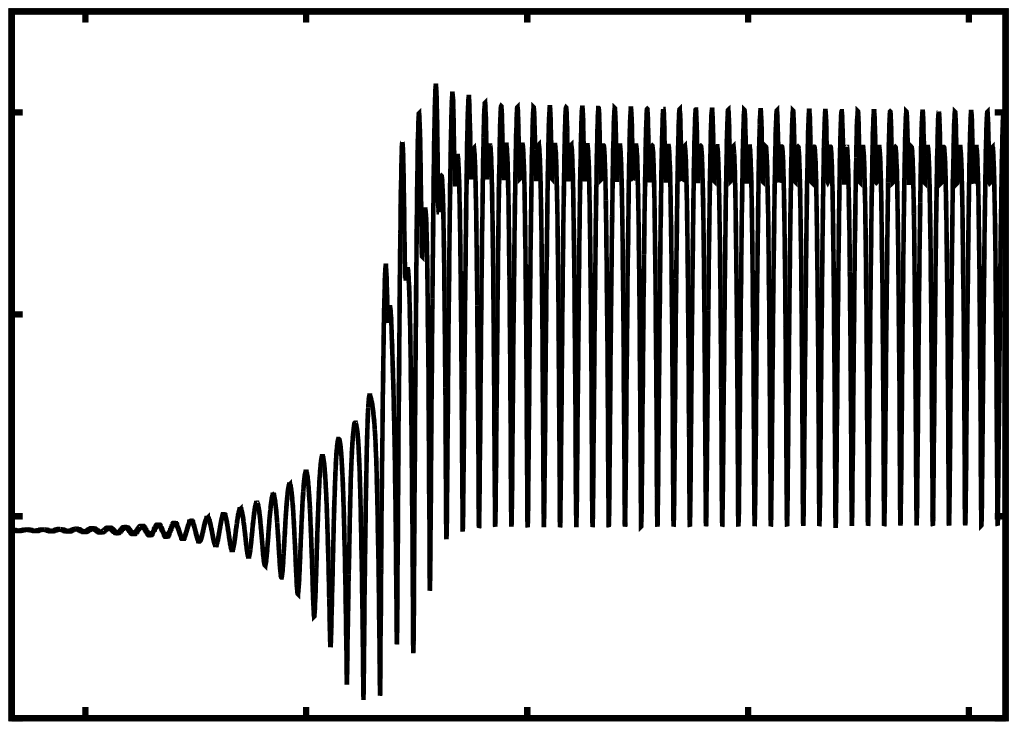} } 
   \scalebox{0.455}{ \input{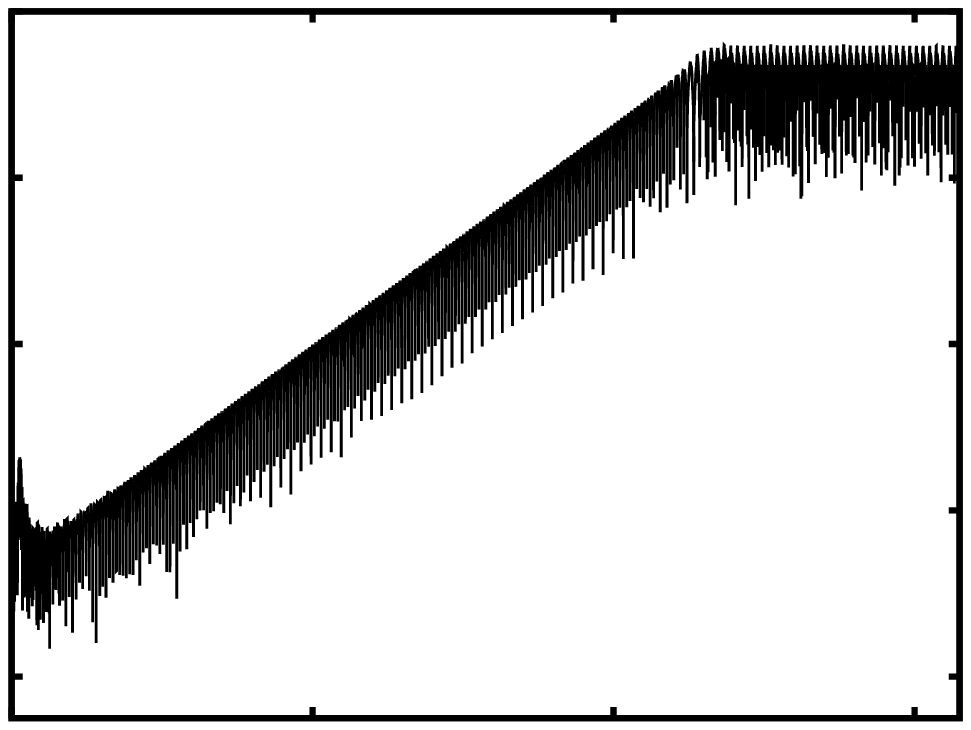} }
  \scalebox{0.455}{ \input{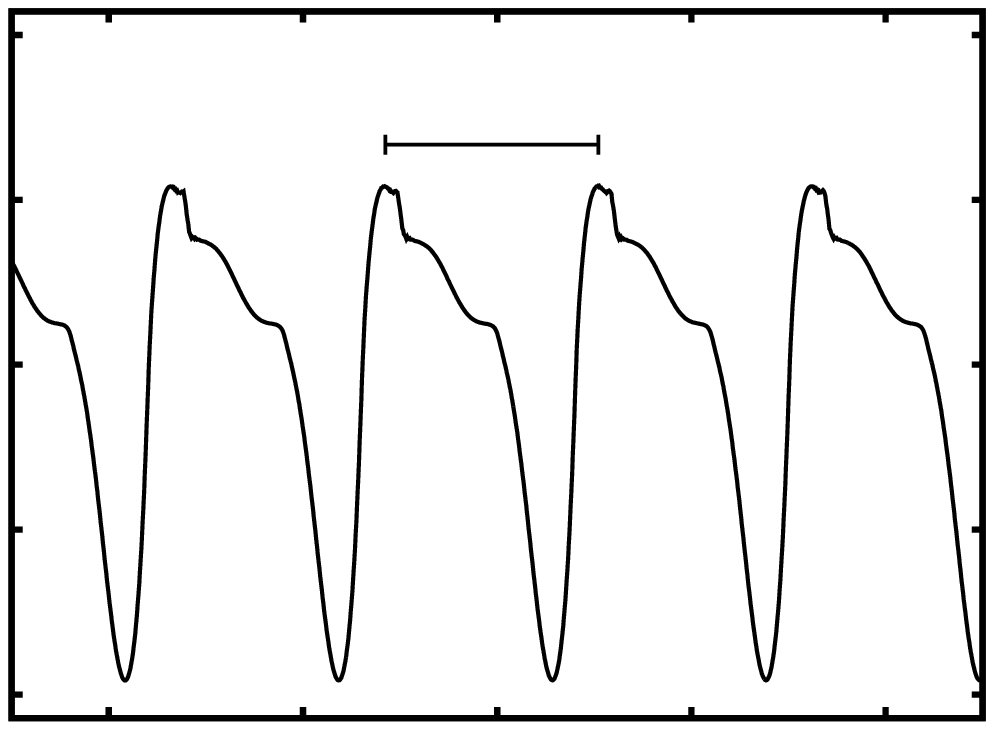} } \\
   \end{array}$
 \caption{Evolution of an instability into the non-linear regime for a model of $\kappa$-Cassiopeiae (HD\,2905) with 
 a mass of 38 M$_{\sun}$: Radius (a) and absolute velocity (b) at the outermost grid point and variations of 
 the bolometric magnitude (c) are given as a function of time.
 This model finally exhibits finite amplitude pulsations with a period of 2.2 days.}
 \normalsize 
 \label{38m_nonlin}
 \end{figure*}

 \begin{figure*}
\centering $
\LARGE
\begin{array}{ccc}
  \scalebox{0.455}{ \input{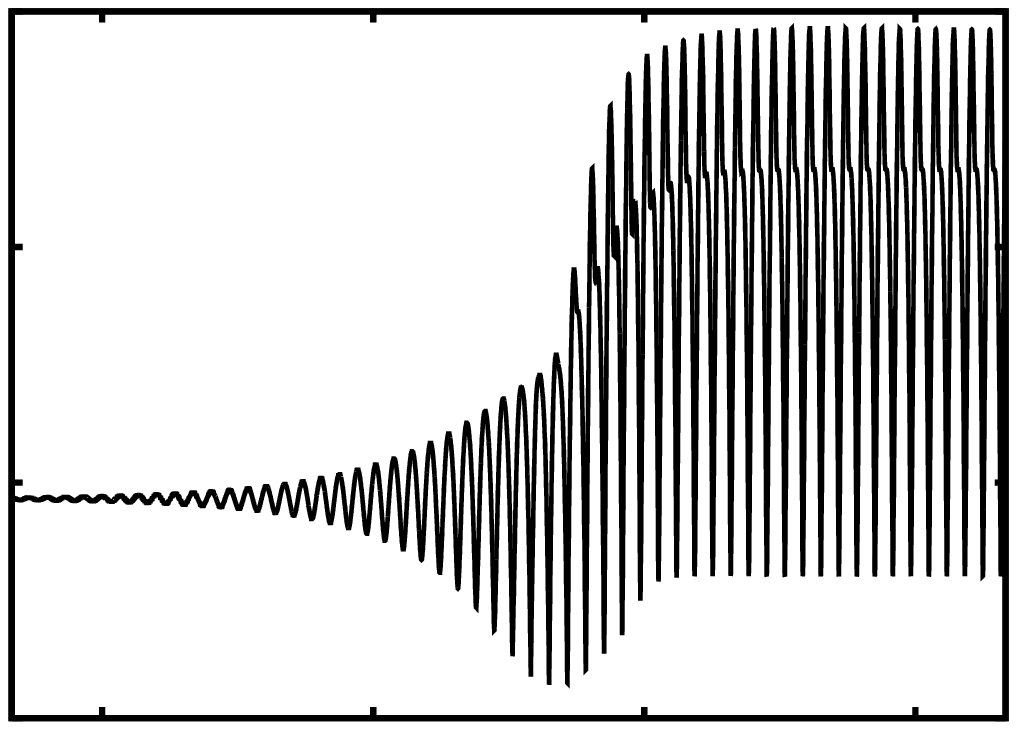} } 
   \scalebox{0.455}{ \input{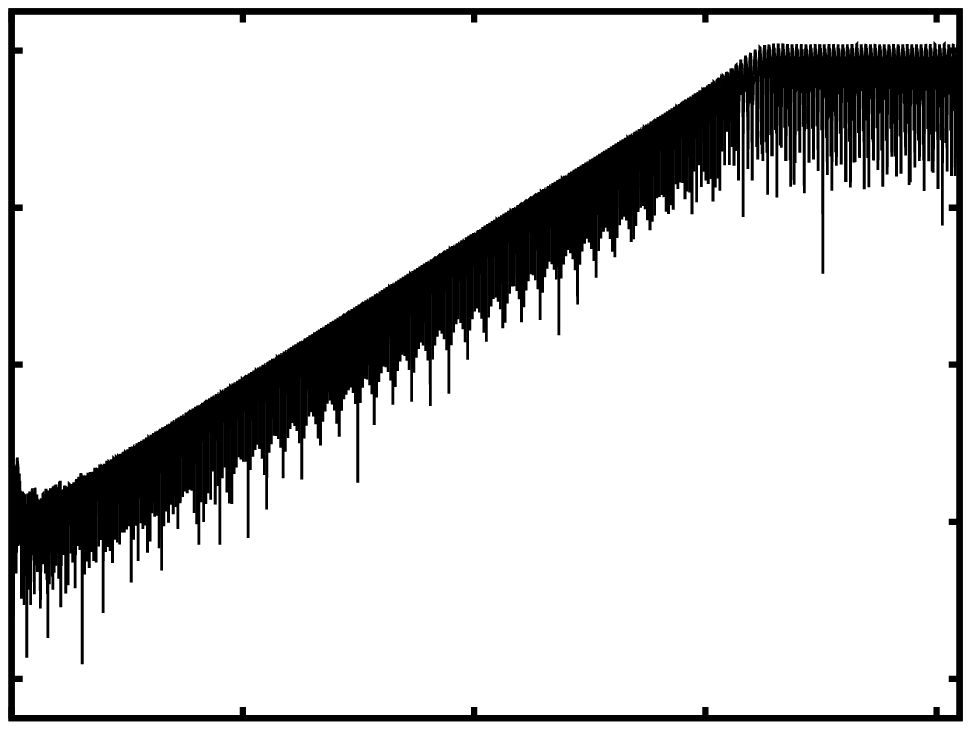} }
  \scalebox{0.455}{ \input{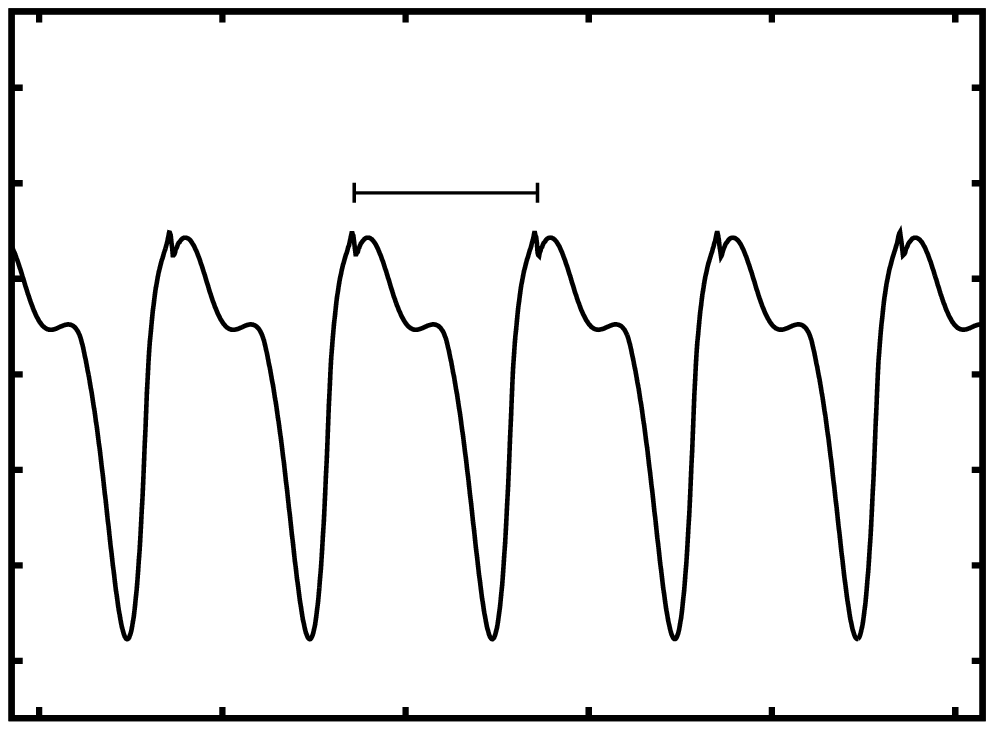} } \\
   \end{array}$
 \caption{Same as Fig.\,\ref{38m_nonlin} but for a model with a mass of 40 M$_{\sun}$. 
 Here the instability leads to finite amplitude pulsations with a period of 2 days.}
 \normalsize 
 \label{40m_nonlin}
 \end{figure*}

 \begin{figure*}
\centering $
\LARGE
\begin{array}{ccc}
  \scalebox{0.455}{ \input{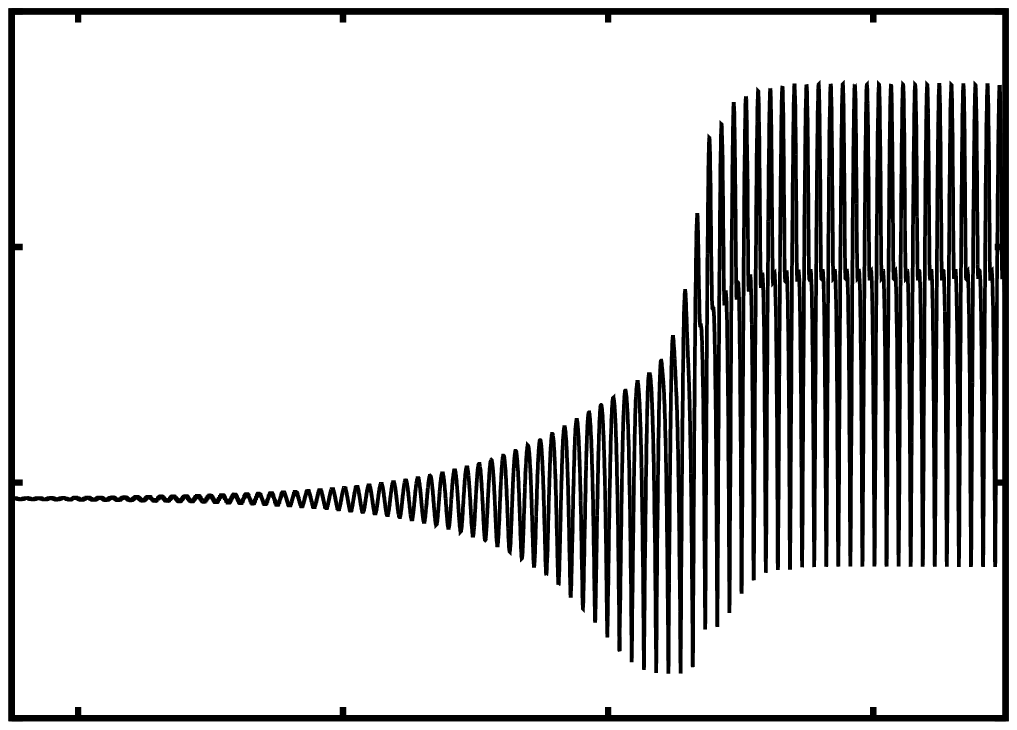} } 
   \scalebox{0.455}{ \input{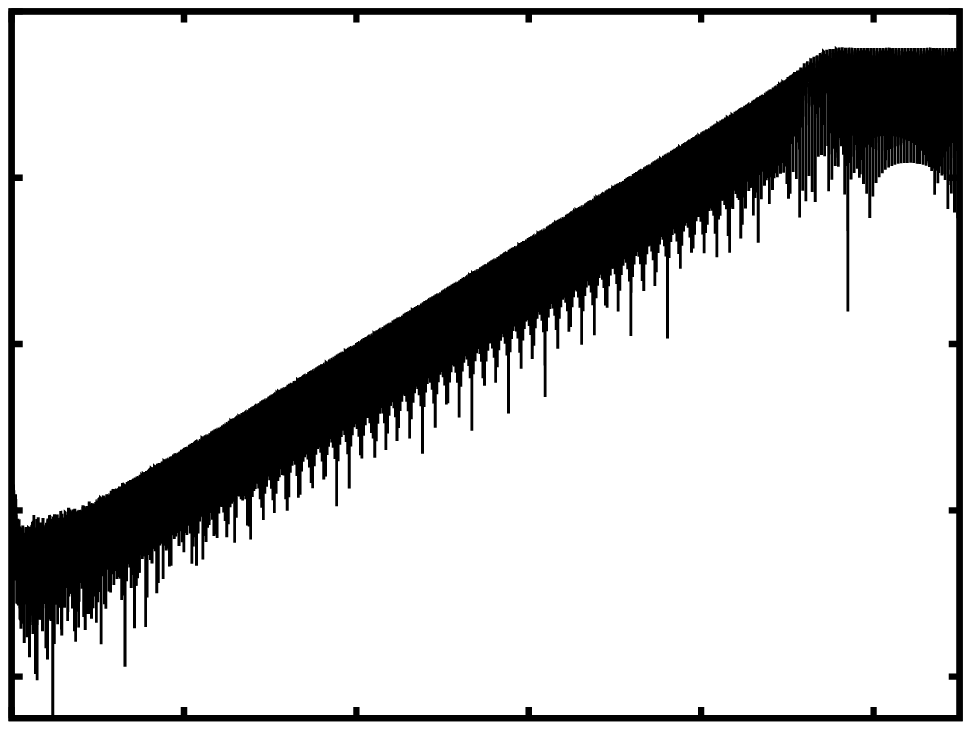} }
  \scalebox{0.455}{ \input{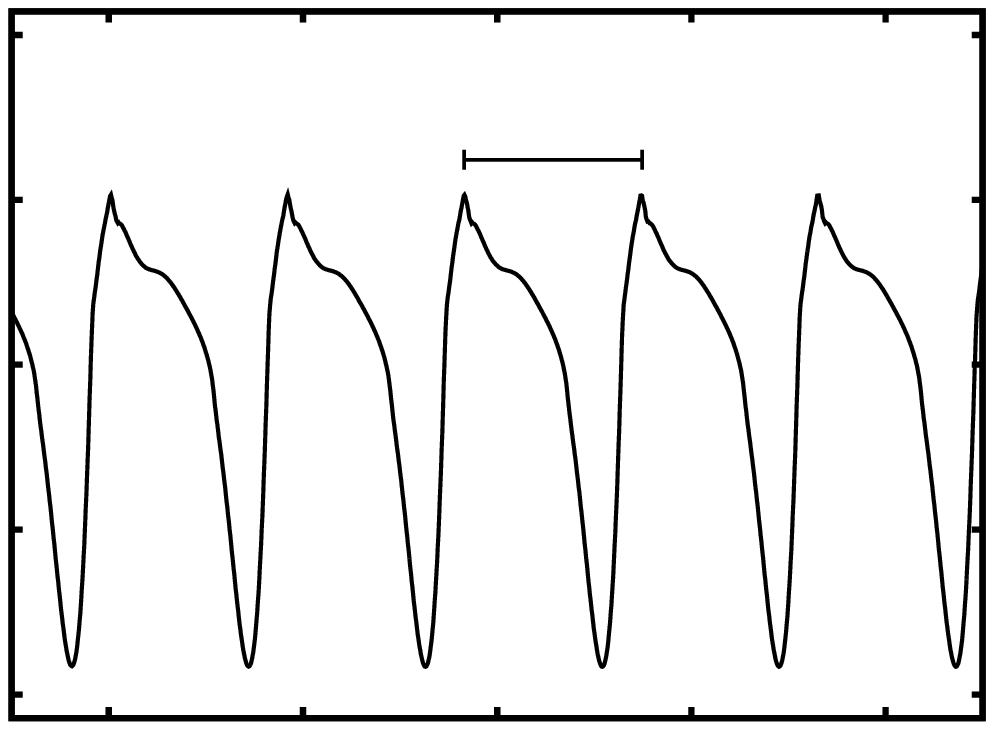} } \\
   \end{array}$
 \caption{Same as Fig.\,\ref{38m_nonlin} but for a model with a mass of 42 M$_{\sun}$. 
 Here the instability leads to finite amplitude pulsations with a period of 1.8 days.}
 \normalsize 
 \label{42m_nonlin}
 \end{figure*}

\subsection{Results of the non-linear simulations}

All models for $\kappa$\,Cassiopeiae with masses in the range between 27 and 44 M$_{\sun}$ are unstable
according to the linear stability analysis. 
In order to determine the final fate of the unstable models, 
we have followed the instabilities into the non-linear regime for selected cases. The results of the non-linear simulation 
for a model with a mass of 
27 M$_{\sun}$ are shown in Fig. \ref{27m_nonlin}: Radius (a), temperature (b) and 
absolute velocity (c) at the the outermost grid point are given as a function of time. 
The evolution of the velocity (starting from hydrostatic equilibrium) shows that the code picks 
up the instability from numerical noise with an amplitude of the order of 10$^{-4}$ cm s$^{-1}$, the
subsequent linear phase is characterized by oscillatory exponential growth.
A saturation of the velocity amplitude is found, when the non-linear regime is reached. 
In this phase the envelope is inflated (see Fig. \ref{27m_nonlin}a) and the radius
of the outermost grid point increases approximately by a factor of nine compared to its  
initial hydrostatic value. As a consequence, the temperature at the outermost grid 
point drops to $\approx$ 5000 K and the simulation had to be stopped, since opacity 
data were no longer available. The maximum velocity reached in the non-linear regime
amounts to 224 km$\,$s$^{-1}$ and corresponds to 43 per cent of the escape 
velocity of the model (517 km$\,$s$^{-1}$).
We note that some simulations of instabilities in models for massive B-type stars have provided
evidence for the instability driven maximum velocity to exceed the escape 
velocity of the corresponding models \citep[see][]{glatzel_1999, yadav_2016} which implies direct mass loss 
due to instabilities. With the reservation that the simulation is not yet complete, we find for the 
model of $\kappa$\,Cassiopeiae considered the maximum velocity to remain well below the escape velocity.
However, the substantial inflation of the envelope in the non-linear regime may be taken as another
indication for mass loss.

 The evolution of instabilities into the non-linear regime for a model with a 
 mass of 32 M$_{\sun}$ is illustrated in  Fig.\,\ref{32m_nonlin}: Radius (a), absolute velocity (b) and temperature (c) at the outermost
 grid point together with the variation of the bolometric magnitude (d) are presented as a function of time.  
 The velocity of the outermost grid point starts from numerical noise in hydrostatic equilibrium and 
 finally saturates after the linear phase of exponential growth
 in the non-linear regime 80 days after start with a maximum velocity close to 126 km s$^{-1}$. 
 Thus finite amplitude pulsations are the result of the instability in this model, which may
 also be deduced from the evolution of the radius (Fig.\,\ref{32m_nonlin}a).
 The final mean value of the radius is approximately 30 per cent larger than the initial hydrostatic value and 
 the inflation of the radius is associated with a decrease of the temperature (Fig.\,\ref{32m_nonlin}c). 
 We emphasize that the temperature at the outermost grid point referred to here is not necessarily related to 
 the effective temperature of the model, since the relative position of the photosphere may differ 
 from its initial hydrostatic location, in particular, when inflation is significant. Moreover, it 
 may vary during the pulsation cycle. 
 The variation of the bolometric magnitude (d) exhibits a well defined period of 3 days.

Similar to the model with a mass of 32 M$_{\sun}$, we have followed the instability into the non-linear 
regime for the model with a mass of 34.5 M$_{\sun}$. The results for this model presented in Fig.\,\ref{345m_nonlin} 
are similar to those obtained for 32 M$_{\sun}$. The velocity saturates with a maximum value of 110 km\,s$^{-1}$ in the 
non-linear regime and the inflation of the radius (Fig.\,\ref{345m_nonlin}a) is less pronounced than 
for 32 M$_{\sun}$. A period of 2.7 days for the final finite amplitude pulsations is obtained from the 
variation of the bolometric magnitude (c). It agrees with the observed dominant period of 
$\kappa$-Cassiopeiae (HD\,2905) reported by \citet{ssd_2018}.

Quantities associated with outermost grid point are generally used to study envelope inflation, 
pulsation period and energy budget during simulation in nonlinear regime. However, quantities linked with outermost grid 
point may not represent the photospheric values.
Therefore for model with mass 34.5 M$_{\sun}$, radius (a), absolute velocity (b), temperature (c) and density (d)
associated with grid points near the photosphere are given in Fig. \ref{345m_nonlinP}. To identify the location 
of photospheric grid point in our simulations, we have used Stefan-Boltzmann relation. 
The outermost grid point where total flux is equal to $\sigma$\,T$^{4}$ (here $\sigma$   
 and T are Boltzmann constant and temperature, respectively) has been taken as an approximate location of the photosphere.
As found 
in Fig.\,\ref{345m_nonlin}, a period of 2.7 days is also present in the variation profile of the mentioned quantities in Fig.\,\ref{345m_nonlinP}. 
Compared to the hydrostatic value of radius (see Fig.\,\ref{345m_nonlin}a), the model exhibits finite amplitude pulsation with 
slightly inflated mean radius ($\approx$ 2.87 $\times$ 10$^{12}$ cm). 
The maximum-to-minimum variation in bolometric magnitude is found to be approximately 0.2 mag which is larger than the value determined 
by \citet{ssd_2018} using HIPPARCOS data. 

Apart from 2.7 days variability, \citet{ssd_2018} has found additional frequencies
in the range of 0.1 and 0.4 d$^{-1}$. Comparison of the shape of the observed bolometric magnitude profile with theoretically
determined bolometric magnitude profile has to be done with caution as the latter is resulting from a single instability 
corresponding to a linear period of 2.7 days while several frequencies are present in the observed profile.
Maximum-to-minimum variation of absolute velocity associated with this grid point (b) is around 240 km\,s$^{-1}$ which is considerably
larger than the values (30 km\,s$^{-1}$) determined by \citet{ssd_2018} using photospheric spectral lines such as Si III and Si IV.
These authors have also reported that the amplitude of variability increases from deep photospheric lines to the lines 
formed in stellar winds. During the finite amplitude pulsation phase, temperature varies in the range of 26820 K and 17500 K 
with variation mostly between 17800\,K and 20000\,K (see Fig.\,\ref{345m_nonlinP}c). From spectroscopic 
analysis, \citet{ssd_2018} find surface temperature in the range of 23700\,K and 25100\,K with average value of 24600 $\pm$ 300\,K. The value of 
artificial viscosity parameter can affect the temperature variation during the finite amplitude pulsation as shown by \cite{yadav_2017b}.
Variation profile of density (d) exhibits sharp peaks with a separation 2.7 days.

Results of the numerical simulation for models with masses of 38, 40 and 42 M$_{\sun}$ 
are presented in Figs.  
\ref{38m_nonlin}, \ref{40m_nonlin} and \ref{42m_nonlin}, respectively. In these figures, the radius (a) 
and the absolute absolute velocity (b) at the outermost grid point 
together with the variation of the bolometric magnitude (c) are given as a function of time. 
The velocity starts from numerical noise at the level of 10$^{-4}$ cm\,s$^{-1}$
and finally saturates in the non-linear regime with maximum values of 
95, 109 and 108 km\,s$^{-1}$ for the models with a mass
of 38, 40, and 42 M$_{\sun}$, respectively. Thus the instabilities
of models with a mass of 38, 40 and 42 M$_{\sun}$ lead to finite amplitude pulsations with 
periods of 2.2, 2.0 and 1.8 days
respectively (see Figs. \ref{38m_nonlin}c, \ref{40m_nonlin}c and \ref{42m_nonlin}c).
From Figs. \ref{38m_nonlin}(a), \ref{40m_nonlin}(a) and \ref{42m_nonlin}(a) we still deduce a 
slight inflation of the radius in the non-linear regime.

% ================ Section 5 ===============
% ================ Section 5 ===============

\section{Discussion and Conclusion}

In order to understand the variabilities observed in $\kappa$\,Cassiopeiae \citep[see e.g.][]{ssd_2018}, we have 
performed a linear non-adiabatic 
stability analysis for models of this supergiant. All models considered with masses in the range between 27 and 44 M$_{\sun}$
are unstable. The instabilities are associated with low order modes (fundamental mode and first overtone) 
while higher order modes
are damped (see Fig. \ref{modal_1}). As expected, growth rate and strength of the instabilities increase with the luminosity to 
mass ratio. If the 
luminosity to mass ratio exceeds 10$^{4}$ in solar units (which holds for all models considered here) dynamical instabilities
are to be expected \citep[see e.g.][]{glatzel_1994}. In particular, strange modes and associated instabilities are typically found in stellar 
models with luminosity to mass
ratios in excess of 10$^{4}$ (solar units). In general, strange modes are related to mode coupling phenomena appearing 
as instability bands and avoided crossings. Prominent examples are   
models of massive ZAMS stars \citep{yadav_2017}, Wolf Rayet stars \citep{kiriakidis_1996} and HdC stars \citep{gautschy_1990b}. 
In modal diagrams of models for $\kappa$\,Cassiopeiae
(see, e.g., Fig. \ref{modal_1}), mode coupling phenomena can be identified for higher order modes. However, they
are not associated with instability bands. On the other hand, mode coupling effects do not seem to be involved  
in the unstable low order modes which appear to be ordinary low order p-modes. 
However, to identify the physical origin of the instabilities found here as driven by resonances (mode coupling, strange modes)
or by the $\kappa$-mechanism requires application of the non-adiabatic reversible (NAR) approximation \citep{gautschy_1990b}.
We intend to present a corresponding analysis in a forthcoming paper.
The stability analysis was repeated with different boundary conditions (see Fig. \ref{modal_2}) to test their influence 
on the results. In fact, for models of $\kappa$\,Cassiopeiae the choice of boundary conditions is largely irrelevant, at least as 
long as the instabilities and the range of unstable models are concerned. 
The pulsation frequency of the unstable fundamental mode lies between 3.9 and 1.6 days for
masses of the stellar model between 27 and 44 M$_{\sun}$.

In order to determine the final fate of the unstable models, the instabilities have been followed into the non-linear regime
by numerical simulation for selected cases. Any simulation of stellar instabilities and finite amplitude stellar
pulsations requires an extremely high accuracy which can be achieved only by a fully conservative numerical scheme.
In particular, the energy balance needs to be correct to a very high degree of precision.
We have adopted a numerical scheme which meets all these requirements \citep[for details see e.g.][]{yadav_2018, grott_2005}. 
In order to ensure that the code picks up and follows the physical instabilities, it was validated by
comparing periods and growth rates of the instabilities appearing in the simulation (in the linear
phase of exponential growth) with the independently predetermined values provided by the linear stability
analysis (see section \ref{validation}). 
For models with masses below 29 M$_{\sun}$, the instabilities lead to a substantial inflation of the 
radius (see Fig. \ref{27m_nonlin}). For higher masses, finite amplitude pulsations with periods between 1 and 3 days are 
the final fate of the unstable models considered. Ground and space based observations of $\kappa$\,Cassiopeiae, \citet{ssd_2018} have
revealed a variability with a dominant period of 2.7 days for this object. 
On the other hand, finite amplitude pulsations with a period of 2.7 days are found as the final result 
when simulating the evolution of the instability into the non-linear regime of the model for $\kappa$\,Cassiopeiae
with a mass of 34.5 M$_{\sun}$. Thus the observed  
variability with a period of 2.7 days may be explained by the instability of the fundamental radial mode
of an object with a mass of 34.5 M$_{\sun}$. 
In addition to the 2.7 day period, the variability of $\kappa$\,Cassiopeiae exhibits additional
periods in the range between 2.5 and 10 days \citep{ssd_2018}. 
According to our linear stability analysis the model with the lowest mass (27 M$_{\sun}$) provides two unstable modes 
with periods of 2.3 and 3.9 days respectively.
Thus both from the linear analysis and the non-linear simulations we conclude that variabilities 
with periods above 3.9 days cannot be explained by 
radial unstable modes of $\kappa$\,Cassiopeiae. Variabilities with periods above 3.9 days might be due to 
gravity modes. Their study requires a linear non-adiabatic stability analysis with respect to
non-radial perturbations which will be the primary objective of an extension of the present study. 
Earlier studies of $\kappa$\,Cassiopeiae report on variabilities on timescales between
1 and 2 hours \cite[see e.g.,][]{elst_1979, badalia_1982}. However,
variability with short periods (of the order of hours) have not been found by \citet{ssd_2018}.
Thus the existence of long-lived short period variabilities
in $\kappa$\,Cassiopeiae is yet to be confirmed. In this context, observations of TESS may contribute 
significantly to our understanding of the 
variabilities in $\kappa$ Cassiopeiae.

From the present study which has been restricted to considering radial perturbations 
we infer that the dominant variability of $\kappa$\,Cassiopeiae with a
period of 2.7 days may be understood as a radial mode pulsation. However, we note that the theoretical light 
curve is differing from the observed light curve of \citet{ssd_2018}. Shape and amplitude of the theoretical light 
curve depends on parameters including the number of unstable modes, strength of present instabilities and value 
of artificial viscosity parameter. Present study is a first step to understand the variability of 2.7 days 
using linear stability analysis in combination with nonlinear simulations. To compare the light curve, we 
need to perform extensive nonlinear simulations. Additional observed variabilities of this star need further 
attention. In particular, variabilities with periods above 10 days being compatible with g-mode pulsations have been observed in 
$\kappa$\,Cassiopeiae \citep{ssd_2018}. 
Moreover, in addition to radial instabilities, non-radial instabilities have been identified in models of B-type stars 
\citep[e.g.][]{guatschy_1993, glatzel_1996, saio_2006, saio_2011}. 
Therefore, to complete and improve our understanding of variabilities in $\kappa$ Cassiopeiae
a linear non-adiabatic stability analysis with respect to non-radial perturbations is inevitable. 
These issues will be presented in a forthcoming paper. 

\section*{Acknowledgements}

APY gratefully acknowledges the hospitality and support of the Aryabhatta Research Institute of Observational Sciences (ARIES) 
Nainital where a part of this study has been performed. We thank the anonymous referee for constructive comments to improve this paper.  

%%%%%%%%%%%%%%%%%%%%%%%%%%%%%%%%%%%%%%%%%%%%%%%%%% 



\bibliographystyle{mnras}
\bibliography{first} % if your bibtex file is called example.bib

\begin{thebibliography}{}
\makeatletter
\relax
\def\mn@urlcharsother{\let\do\@makeother \do\$\do\&\do\#\do\^\do\_\do\%\do\~}
\def\mn@doi{\begingroup\mn@urlcharsother \@ifnextchar [ {\mn@doi@}
  {\mn@doi@[]}}
\def\mn@doi@[#1]#2{\def\@tempa{#1}\ifx\@tempa\@empty \href
  {http://dx.doi.org/#2} {doi:#2}\else \href {http://dx.doi.org/#2} {#1}\fi
  \endgroup}
\def\mn@eprint#1#2{\mn@eprint@#1:#2::\@nil}
\def\mn@eprint@arXiv#1{\href {http://arxiv.org/abs/#1} {{\tt arXiv:#1}}}
\def\mn@eprint@dblp#1{\href {http://dblp.uni-trier.de/rec/bibtex/#1.xml}
  {dblp:#1}}
\def\mn@eprint@#1:#2:#3:#4\@nil{\def\@tempa {#1}\def\@tempb {#2}\def\@tempc
  {#3}\ifx \@tempc \@empty \let \@tempc \@tempb \let \@tempb \@tempa \fi \ifx
  \@tempb \@empty \def\@tempb {arXiv}\fi \@ifundefined
  {mn@eprint@\@tempb}{\@tempb:\@tempc}{\expandafter \expandafter \csname
  mn@eprint@\@tempb\endcsname \expandafter{\@tempc}}}

\bibitem[\protect\citeauthoryear{{Aerts} et~al.,}{{Aerts}
  et~al.}{2010}]{aerts_2010arti}
{Aerts} C.,  et~al., 2010, \mn@doi [\aap] {10.1051/0004-6361/201014124}, \href
  {http://adsabs.harvard.edu/abs/2010A%26A...513L..11A} {513, L11}

\bibitem[\protect\citeauthoryear{{Badalia} \& {Gurm}}{{Badalia} \&
  {Gurm}}{1982}]{badalia_1982}
{Badalia} J.~K.,  {Gurm} H.~S.,  1982, Information Bulletin on Variable Stars,
  \href {http://cdsads.u-strasbg.fr/abs/1982IBVS.2163....1B} {2163}

\bibitem[\protect\citeauthoryear{{Baker} \& {Kippenhahn}}{{Baker} \&
  {Kippenhahn}}{1965}]{baker_1965}
{Baker} N.,  {Kippenhahn} R.,  1965, \mn@doi [\apj] {10.1086/148359}, \href
  {http://adsabs.harvard.edu/abs/1965ApJ...142..868B} {142, 868}

\bibitem[\protect\citeauthoryear{{Balona} et~al.,}{{Balona}
  et~al.}{2011}]{balona_2011}
{Balona} L.~A.,  et~al., 2011, \mn@doi [\mnras]
  {10.1111/j.1365-2966.2011.18311.x}, \href
  {http://cdsads.u-strasbg.fr/abs/2011MNRAS.413.2403B} {413, 2403}

\bibitem[\protect\citeauthoryear{{B{\"o}hm-Vitense}}{{B{\"o}hm-Vitense}}{1958}]{bohm_1958}
{B{\"o}hm-Vitense} E.,  1958, \zap, \href
  {http://adsabs.harvard.edu/abs/1958ZA.....46..108B} {46, 108}

\bibitem[\protect\citeauthoryear{{Elst}}{{Elst}}{1979}]{elst_1979}
{Elst} E.~W.,  1979, Information Bulletin on Variable Stars, \href
  {http://adsabs.harvard.edu/abs/1979IBVS.1697....1E} {1697}

\bibitem[\protect\citeauthoryear{{Evans}, {Crowther}, {Fullerton}  \&
  {Hillier}}{{Evans} et~al.}{2004}]{evans_2004}
{Evans} C.~J.,  {Crowther} P.~A.,  {Fullerton} A.~W.,   {Hillier} D.~J.,  2004,
  \mn@doi [\apj] {10.1086/421769}, \href
  {http://cdsads.u-strasbg.fr/abs/2004ApJ...610.1021E} {610, 1021}

\bibitem[\protect\citeauthoryear{{Gautschy} \& {Glatzel}}{{Gautschy} \&
  {Glatzel}}{1990a}]{gautschy_1990a}
{Gautschy} A.,  {Glatzel} W.,  1990a, \mnras, \href
  {http://adsabs.harvard.edu/abs/1990MNRAS.245..154G} {245, 154}

\bibitem[\protect\citeauthoryear{{Gautschy} \& {Glatzel}}{{Gautschy} \&
  {Glatzel}}{1990b}]{gautschy_1990b}
{Gautschy} A.,  {Glatzel} W.,  1990b, \mnras, \href
  {http://adsabs.harvard.edu/abs/1990MNRAS.245..597G} {245, 597}

\bibitem[\protect\citeauthoryear{{Gautschy} \& {Saio}}{{Gautschy} \&
  {Saio}}{1993}]{guatschy_1993}
{Gautschy} A.,  {Saio} H.,  1993, \mn@doi [\mnras] {10.1093/mnras/262.1.213},
  \href {https://ui.adsabs.harvard.edu/abs/1993MNRAS.262..213G} {262, 213}

\bibitem[\protect\citeauthoryear{{Glatzel}}{{Glatzel}}{1994}]{glatzel_1994}
{Glatzel} W.,  1994, \mnras, \href
  {http://adsabs.harvard.edu/abs/1994MNRAS.271...66G} {271, 66}

\bibitem[\protect\citeauthoryear{{Glatzel} \& {Chernigovski}}{{Glatzel} \&
  {Chernigovski}}{2016}]{glatzel_2016}
{Glatzel} W.,  {Chernigovski} S.,  2016, \mn@doi [\mnras]
  {10.1093/mnras/stw003}, \href
  {http://adsabs.harvard.edu/abs/2016MNRAS.457.1190G} {457, 1190}

\bibitem[\protect\citeauthoryear{{Glatzel} \& {Kiriakidis}}{{Glatzel} \&
  {Kiriakidis}}{1993}]{glatzel_1993}
{Glatzel} W.,  {Kiriakidis} M.,  1993, \mnras, \href
  {http://adsabs.harvard.edu/abs/1993MNRAS.262...85G} {262, 85}

\bibitem[\protect\citeauthoryear{{Glatzel} \& {Mehren}}{{Glatzel} \&
  {Mehren}}{1996}]{glatzel_1996}
{Glatzel} W.,  {Mehren} S.,  1996, \mn@doi [\mnras] {10.1093/mnras/282.4.1470},
  \href {http://adsabs.harvard.edu/abs/1996MNRAS.282.1470G} {282, 1470}

\bibitem[\protect\citeauthoryear{{Glatzel}, {Kiriakidis}  \&
  {Fricke}}{{Glatzel} et~al.}{1993}]{glatzel_1993b}
{Glatzel} W.,  {Kiriakidis} M.,   {Fricke} K.~J.,  1993, \mnras, \href
  {http://adsabs.harvard.edu/abs/1993MNRAS.262L...7G} {262, L7}

\bibitem[\protect\citeauthoryear{{Glatzel}, {Kiriakidis}, {Chernigovskij}  \&
  {Fricke}}{{Glatzel} et~al.}{1999}]{glatzel_1999}
{Glatzel} W.,  {Kiriakidis} M.,  {Chernigovskij} S.,   {Fricke} K.~J.,  1999,
  \mn@doi [\mnras] {10.1046/j.1365-8711.1999.02190.x}, \href
  {http://adsabs.harvard.edu/abs/1999MNRAS.303..116G} {303, 116}

\bibitem[\protect\citeauthoryear{{Grott}, {Chernigovski}  \& {Glatzel}}{{Grott}
  et~al.}{2005}]{grott_2005}
{Grott} M.,  {Chernigovski} S.,   {Glatzel} W.,  2005, \mn@doi [\mnras]
  {10.1111/j.1365-2966.2005.09162.x}, \href
  {http://adsabs.harvard.edu/abs/2005MNRAS.360.1532G} {360, 1532}

\bibitem[\protect\citeauthoryear{{Gvaramadze}, {Kniazev}, {Kroupa}  \&
  {Oh}}{{Gvaramadze} et~al.}{2011}]{gvaramadze_2011}
{Gvaramadze} V.~V.,  {Kniazev} A.~Y.,  {Kroupa} P.,   {Oh} S.,  2011, \mn@doi
  [\aap] {10.1051/0004-6361/201117746}, \href
  {https://ui.adsabs.harvard.edu/abs/2011A%26A...535A..29G} {535, A29}

\bibitem[\protect\citeauthoryear{{Haucke}, {Cidale}, {Venero}, {Cur{\'e}},
  {Kraus}, {Kanaan}  \& {Arcos}}{{Haucke} et~al.}{2018}]{haucke_2018}
{Haucke} M.,  {Cidale} L.~S.,  {Venero} R.~O.~J.,  {Cur{\'e}} M.,  {Kraus} M.,
  {Kanaan} S.,   {Arcos} C.,  2018, \mn@doi [\aap]
  {10.1051/0004-6361/201731678}, \href
  {https://ui.adsabs.harvard.edu/abs/2018A%26A...614A..91H} {614, A91}

\bibitem[\protect\citeauthoryear{{Hayes}}{{Hayes}}{1984}]{hayes_1984}
{Hayes} D.~P.,  1984, \mn@doi [\aj] {10.1086/113616}, \href
  {http://cdsads.u-strasbg.fr/abs/1984AJ.....89.1219H} {89, 1219}

\bibitem[\protect\citeauthoryear{{Iglesias} \& {Rogers}}{{Iglesias} \&
  {Rogers}}{1996}]{iglesias_1996}
{Iglesias} C.~A.,  {Rogers} F.~J.,  1996, \mn@doi [\apj] {10.1086/177381},
  \href {http://adsabs.harvard.edu/abs/1996ApJ...464..943I} {464, 943}

\bibitem[\protect\citeauthoryear{{Katushkina}, {Alexashov}, {Gvaramadze}  \&
  {Izmodenov}}{{Katushkina} et~al.}{2018}]{katushkina_2018}
{Katushkina} O.~A.,  {Alexashov} D.~B.,  {Gvaramadze} V.~V.,   {Izmodenov}
  V.~V.,  2018, \mn@doi [\mnras] {10.1093/mnras/stx2488}, \href
  {http://cdsads.u-strasbg.fr/abs/2018MNRAS.473.1576K} {473, 1576}

\bibitem[\protect\citeauthoryear{{Kippenhahn}, {Weigert}  \&
  {Weiss}}{{Kippenhahn} et~al.}{2012}]{kippenhahn_2012}
{Kippenhahn} R.,  {Weigert} A.,   {Weiss} A.,  2012, {Stellar Structure and
  Evolution}, \mn@doi{10.1007/978-3-642-30304-3.
}

\bibitem[\protect\citeauthoryear{{Kiriakidis}, {Glatzel}  \&
  {Fricke}}{{Kiriakidis} et~al.}{1996}]{kiriakidis_1996}
{Kiriakidis} M.,  {Glatzel} W.,   {Fricke} K.~J.,  1996, \mn@doi [\mnras]
  {10.1093/mnras/281.2.406}, \href
  {http://cdsads.u-strasbg.fr/abs/1996MNRAS.281..406K} {281, 406}

\bibitem[\protect\citeauthoryear{{Koen} \& {Eyer}}{{Koen} \&
  {Eyer}}{2002}]{koen_2002}
{Koen} C.,  {Eyer} L.,  2002, \mn@doi [\mnras]
  {10.1046/j.1365-8711.2002.05150.x}, \href
  {http://adsabs.harvard.edu/abs/2002MNRAS.331...45K} {331, 45}

\bibitem[\protect\citeauthoryear{{Kraus} et~al.,}{{Kraus}
  et~al.}{2015}]{kraus_2015}
{Kraus} M.,  et~al., 2015, \mn@doi [\aap] {10.1051/0004-6361/201425383}, \href
  {http://cdsads.u-strasbg.fr/abs/2015A%26A...581A..75K} {581, A75}

\bibitem[\protect\citeauthoryear{{Kudritzki}, {Puls}, {Lennon}, {Venn},
  {Reetz}, {Najarro}, {McCarthy}  \& {Herrero}}{{Kudritzki}
  et~al.}{1999}]{kudritzki_1999}
{Kudritzki} R.~P.,  {Puls} J.,  {Lennon} D.~J.,  {Venn} K.~A.,  {Reetz} J.,
  {Najarro} F.,  {McCarthy} J.~K.,   {Herrero} A.,  1999, \aap, \href
  {http://cdsads.u-strasbg.fr/abs/1999A%26A...350..970K} {350, 970}

\bibitem[\protect\citeauthoryear{{Percy}}{{Percy}}{1981}]{percy_1981}
{Percy} J.~R.,  1981, Information Bulletin on Variable Stars, \href
  {http://cdsads.u-strasbg.fr/abs/1981IBVS.1946....1P} {1946}

\bibitem[\protect\citeauthoryear{{Rivinius}, {Baade}  \& {Carciofi}}{{Rivinius}
  et~al.}{2016}]{rivinius_2016}
{Rivinius} T.,  {Baade} D.,   {Carciofi} A.~C.,  2016, \mn@doi [\aap]
  {10.1051/0004-6361/201628411}, \href
  {http://cdsads.u-strasbg.fr/abs/2016A%26A...593A.106R} {593, A106}

\bibitem[\protect\citeauthoryear{{Rogers} \& {Iglesias}}{{Rogers} \&
  {Iglesias}}{1992}]{rogers_1992}
{Rogers} F.~J.,  {Iglesias} C.~A.,  1992, \mn@doi [\apjs] {10.1086/191659},
  \href {http://adsabs.harvard.edu/abs/1992ApJS...79..507R} {79, 507}

\bibitem[\protect\citeauthoryear{{Rogers}, {Swenson}  \& {Iglesias}}{{Rogers}
  et~al.}{1996}]{rogers_1996}
{Rogers} F.~J.,  {Swenson} F.~J.,   {Iglesias} C.~A.,  1996, \mn@doi [\apj]
  {10.1086/176705}, \href {http://adsabs.harvard.edu/abs/1996ApJ...456..902R}
  {456, 902}

\bibitem[\protect\citeauthoryear{{Rybicka}, {Zoc{\l}o{\'n}ska}  \&
  {Tomi{\'c}}}{{Rybicka} et~al.}{2018}]{rybicka_2018}
{Rybicka} M.,  {Zoc{\l}o{\'n}ska} E.,   {Tomi{\'c}} S.,  2018, in {Wade} G.~A.,
   {Baade} D.,  {Guzik} J.~A.,   {Smolec} R.,  eds,  Vol. 8, 3rd BRITE Science
  Conference. pp 134--138

\bibitem[\protect\citeauthoryear{{Saesen}, {Briquet}, {Aerts}, {Miglio}  \&
  {Carrier}}{{Saesen} et~al.}{2013}]{saesen_2013}
{Saesen} S.,  {Briquet} M.,  {Aerts} C.,  {Miglio} A.,   {Carrier} F.,  2013,
  \mn@doi [\aj] {10.1088/0004-6256/146/4/102}, \href
  {http://cdsads.u-strasbg.fr/abs/2013AJ....146..102S} {146, 102}

\bibitem[\protect\citeauthoryear{{Saio}}{{Saio}}{2011}]{saio_2011}
{Saio} H.,  2011, \mn@doi [\mnras] {10.1111/j.1365-2966.2010.18019.x}, \href
  {http://adsabs.harvard.edu/abs/2011MNRAS.412.1814S} {412, 1814}

\bibitem[\protect\citeauthoryear{{Saio} et~al.,}{{Saio}
  et~al.}{2006}]{saio_2006}
{Saio} H.,  et~al., 2006, \mn@doi [\apj] {10.1086/507409}, \href
  {http://cdsads.u-strasbg.fr/abs/2006ApJ...650.1111S} {650, 1111}

\bibitem[\protect\citeauthoryear{{Searle}, {Prinja}, {Massa}  \&
  {Ryans}}{{Searle} et~al.}{2008}]{searle_2008}
{Searle} S.~C.,  {Prinja} R.~K.,  {Massa} D.,   {Ryans} R.,  2008, \mn@doi
  [\aap] {10.1051/0004-6361:20077125}, \href
  {http://adsabs.harvard.edu/abs/2008A%26A...481..777S} {481, 777}

\bibitem[\protect\citeauthoryear{{Sim{\'o}n-D{\'{\i}}az}, {Aerts}, {Urbaneja},
  {Camacho}, {Antoci}, {Fredslund Andersen}, {Grundahl}  \&
  {Pall{\'e}}}{{Sim{\'o}n-D{\'{\i}}az} et~al.}{2018}]{ssd_2018}
{Sim{\'o}n-D{\'{\i}}az} S.,  {Aerts} C.,  {Urbaneja} M.~A.,  {Camacho} I.,
  {Antoci} V.,  {Fredslund Andersen} M.,  {Grundahl} F.,   {Pall{\'e}} P.~L.,
  2018, \mn@doi [\aap] {10.1051/0004-6361/201732160}, \href
  {https://ui.adsabs.harvard.edu/abs/2018A%26A...612A..40S} {612, A40}

\bibitem[\protect\citeauthoryear{{Smartt}, {Lennon}, {Kudritzki}, {Rosales},
  {Ryans}  \& {Wright}}{{Smartt} et~al.}{2002}]{smartt_2002}
{Smartt} S.~J.,  {Lennon} D.~J.,  {Kudritzki} R.~P.,  {Rosales} F.,  {Ryans}
  R.~S.~I.,   {Wright} N.,  2002, \mn@doi [\aap] {10.1051/0004-6361:20020829},
  \href {http://cdsads.u-strasbg.fr/abs/2002A%26A...391..979S} {391, 979}

\bibitem[\protect\citeauthoryear{{Underhill}}{{Underhill}}{1979}]{underhill_1979}
{Underhill} A.~B.,  1979, \mn@doi [\apj] {10.1086/157526}, \href
  {http://cdsads.u-strasbg.fr/abs/1979ApJ...234..528U} {234, 528}

\bibitem[\protect\citeauthoryear{{Waelkens}, {Aerts}, {Kestens}, {Grenon}  \&
  {Eyer}}{{Waelkens} et~al.}{1998}]{waelkens_1998}
{Waelkens} C.,  {Aerts} C.,  {Kestens} E.,  {Grenon} M.,   {Eyer} L.,  1998,
  \aap, \href {https://ui.adsabs.harvard.edu/abs/1998A%26A...330..215W} {330,
  215}

\bibitem[\protect\citeauthoryear{{Yadav} \& {Glatzel}}{{Yadav} \&
  {Glatzel}}{2016}]{yadav_2016}
{Yadav} A.~P.,  {Glatzel} W.,  2016, \mn@doi [\mnras] {10.1093/mnras/stw236},
  \href {http://adsabs.harvard.edu/abs/2016MNRAS.457.4330Y} {457, 4330}

\bibitem[\protect\citeauthoryear{{Yadav} \& {Glatzel}}{{Yadav} \&
  {Glatzel}}{2017a}]{yadav_2017}
{Yadav} A.~P.,  {Glatzel} W.,  2017a, \mn@doi [\mnras] {10.1093/mnras/stw2734},
  \href {http://adsabs.harvard.edu/abs/2017MNRAS.465..234Y} {465, 234}

\bibitem[\protect\citeauthoryear{{Yadav} \& {Glatzel}}{{Yadav} \&
  {Glatzel}}{2017b}]{yadav_2017b}
{Yadav} A.~P.,  {Glatzel} W.,  2017b, \mn@doi [\mnras] {10.1093/mnras/stx1808},
  \href {http://adsabs.harvard.edu/abs/2017MNRAS.471.3245Y} {471, 3245}

\bibitem[\protect\citeauthoryear{{Yadav}, {K{\"u}hnrich Biavatti}  \&
  {Glatzel}}{{Yadav} et~al.}{2018}]{yadav_2018}
{Yadav} A.~P.,  {K{\"u}hnrich Biavatti} S.~H.,   {Glatzel} W.,  2018, \mn@doi
  [\mnras] {10.1093/mnras/sty092}, \href
  {http://cdsads.u-strasbg.fr/abs/2018MNRAS.475.4881Y} {475, 4881}

\bibitem[\protect\citeauthoryear{{van Buren} \& {McCray}}{{van Buren} \&
  {McCray}}{1988}]{buren_1988}
{van Buren} D.,  {McCray} R.,  1988, \mn@doi [\apjl] {10.1086/185184}, \href
  {https://ui.adsabs.harvard.edu/abs/1988ApJ...329L..93V} {329, L93}

\makeatother
\end{thebibliography}

%\begin{thebibliography}{99}

%%%%%%%%%%%%%%%%%%%%%%%%%%%%%%%%%%%%%%%%%%%%%%%%%%

%%%%%%%%%%%%%%%%% APPENDICES %%%%%%%%%%%%%%%%%%%%%

%\appendix



%%%%%%%%%%%%%%%%%%%%%%%%%%%%%%%%%%%%%%%%%%%%%%%%%%

% Don't change these lines
\bsp	% typesetting comment
\label{lastpage}
\end{document}